\documentclass[ALICE,manyauthors]{cernphprep}
\usepackage[comma,square,numbers,sort&compress]{natbib}
\usepackage{hyperref}
\usepackage{lineno}
\usepackage{xspace}
\usepackage{xcolor}
\usepackage[T1]{fontenc}
\usepackage{orcidlink} 

\usepackage{cleveref}
\Crefname{equation}{Equation}{Equations}
\Crefname{figure}{Figure}{Figures}
\Crefname{section}{Section}{Sections}
\Crefname{tabular}{Table}{Tables}

\crefname{equation}{Eq.}{Eqs.}
\crefname{figure}{Fig.}{Figs.}
\crefname{section}{Sec.}{Secs.}
\crefname{tabular}{Tab.}{Tabs.}

\DeclareMathAlphabet{\mathcal}{OMS}{cmsy}{m}{n}

\begin{document}
%%%%%%%%%%%%%%%%%%%%%%%%%%%%%%%%%%%%%%%%%%%%%%%%%%
% These are some new commands that may be useful 
% for paper writing in general. If other newcommands
% are needed for your specific paper, please feel 
% free to add here. 
%
% The currently available commands are organized in: 
% 1) Systems
% 2) Quantities
% 3) Energies and units
% 4) Detectors
% 5) particle species 
%%%%%%%%%%%%%%%%%%%%%%%%%%%%%%%%%%%%%%%%%%%%%%%%%%

% 1) SYSTEMS 
\newcommand{\pp}           {pp\xspace}
\newcommand{\ppbar}        {\mbox{$\mathrm {p\overline{p}}$}\xspace}
\newcommand{\XeXe}         {\mbox{Xe--Xe}\xspace}
\newcommand{\PbPb}         {\mbox{Pb--Pb}\xspace}
\newcommand{\pA}           {\mbox{pA}\xspace}
\newcommand{\pPb}          {\mbox{p--Pb}\xspace}
\newcommand{\AuAu}         {\mbox{Au--Au}\xspace}
\newcommand{\dAu}          {\mbox{d--Au}\xspace}

% 2) QUANTITIES 
\newcommand{\s}            {\ensuremath{\sqrt{s}}\xspace}
\newcommand{\snn}          {\ensuremath{\sqrt{s_{\mathrm{NN}}}}\xspace}
\newcommand{\pt}           {\ensuremath{p_{\rm T}}\xspace}
\newcommand{\meanpt}       {$\langle p_{\mathrm{T}}\rangle$\xspace}
\newcommand{\ycms}         {\ensuremath{y_{\rm CMS}}\xspace}
\newcommand{\ylab}         {\ensuremath{y_{\rm lab}}\xspace}
\newcommand{\etarange}[1]  {\mbox{$\left | \eta \right |~<~#1$}}
\newcommand{\yrange}[1]    {\mbox{$\left | y \right |~<~#1$}}
\newcommand{\dndy}         {\ensuremath{\mathrm{d}N_\mathrm{ch}/\mathrm{d}y}\xspace}
\newcommand{\dndeta}       {\ensuremath{\mathrm{d}N_\mathrm{ch}/\mathrm{d}\eta}\xspace}
\newcommand{\avdndeta}     {\ensuremath{\langle\dndeta\rangle}\xspace}
\newcommand{\dNdy}         {\ensuremath{\mathrm{d}N_\mathrm{ch}/\mathrm{d}y}\xspace}
\newcommand{\Npart}        {\ensuremath{N_\mathrm{part}}\xspace}
\newcommand{\Ncoll}        {\ensuremath{N_\mathrm{coll}}\xspace}
\newcommand{\dEdx}         {\ensuremath{\textrm{d}E/\textrm{d}x}\xspace}
\newcommand{\RpPb}         {\ensuremath{R_{\rm pPb}}\xspace}

% 3) ENERGIES, UNITS
\newcommand{\nineH}        {$\sqrt{s}~=~0.9$~Te\kern-.1emV\xspace}
\newcommand{\seven}        {$\sqrt{s}~=~7$~Te\kern-.1emV\xspace}
\newcommand{\twoH}         {$\sqrt{s}~=~0.2$~Te\kern-.1emV\xspace}
\newcommand{\twosevensix}  {$\sqrt{s}~=~2.76$~Te\kern-.1emV\xspace}
\newcommand{\five}         {$\sqrt{s}~=~5.02$~Te\kern-.1emV\xspace}
\newcommand{\twosevensixnn}{$\sqrt{s_{\mathrm{NN}}}~=~2.76$~Te\kern-.1emV\xspace}
\newcommand{\fivenn}       {$\sqrt{s_{\mathrm{NN}}}~=~5.02$~Te\kern-.1emV\xspace}
\newcommand{\LT}           {L{\'e}vy-Tsallis\xspace}
\newcommand{\GeVc}         {Ge\kern-.1emV/$c$\xspace}
\newcommand{\MeVc}         {Me\kern-.1emV/$c$\xspace}
\newcommand{\TeV}          {Te\kern-.1emV\xspace}
\newcommand{\GeV}          {Ge\kern-.1emV\xspace}
\newcommand{\MeV}          {Me\kern-.1emV\xspace}
\newcommand{\GeVmass}      {Ge\kern-.2emV/$c^2$\xspace}
\newcommand{\MeVmass}      {Me\kern-.2emV/$c^2$\xspace}
\newcommand{\lumi}         {\ensuremath{\mathcal{L}}\xspace}

% 4) DETECTORS 
\newcommand{\ITS}          {\rm{ITS}\xspace}
\newcommand{\TOF}          {\rm{TOF}\xspace}
\newcommand{\ZDC}          {\rm{ZDC}\xspace}
\newcommand{\ZDCs}         {\rm{ZDCs}\xspace}
\newcommand{\ZNA}          {\rm{ZNA}\xspace}
\newcommand{\ZNC}          {\rm{ZNC}\xspace}
\newcommand{\SPD}          {\rm{SPD}\xspace}
\newcommand{\SDD}          {\rm{SDD}\xspace}
\newcommand{\SSD}          {\rm{SSD}\xspace}
\newcommand{\TPC}          {\rm{TPC}\xspace}
\newcommand{\TRD}          {\rm{TRD}\xspace}
\newcommand{\VZERO}        {\rm{V0}\xspace}
\newcommand{\VZEROA}       {\rm{V0A}\xspace}
\newcommand{\VZEROC}       {\rm{V0C}\xspace}
\newcommand{\Vdecay} 	   {\ensuremath{V^{0}}\xspace}

% 4) PARTICLE SPECIES 
\newcommand{\ee}           {\ensuremath{e^{+}e^{-}}} 
\newcommand{\pip}          {\ensuremath{\pi^{+}}\xspace}
\newcommand{\pim}          {\ensuremath{\pi^{-}}\xspace}
\newcommand{\kap}          {\ensuremath{\rm{K}^{+}}\xspace}
\newcommand{\kam}          {\ensuremath{\rm{K}^{-}}\xspace}
\newcommand{\pbar}         {\ensuremath{\rm\overline{p}}\xspace}
\newcommand{\kzero}        {\ensuremath{{\rm K}^{0}_{\rm{S}}}\xspace}
\newcommand{\lmb}          {\ensuremath{\Lambda}\xspace}
\newcommand{\almb}         {\ensuremath{\overline{\Lambda}}\xspace}
\newcommand{\Om}           {\ensuremath{\Omega^-}\xspace}
\newcommand{\Mo}           {\ensuremath{\overline{\Omega}^+}\xspace}
\newcommand{\X}            {\ensuremath{\Xi^-}\xspace}
\newcommand{\Ix}           {\ensuremath{\overline{\Xi}^+}\xspace}
\newcommand{\Xis}          {\ensuremath{\Xi^{\pm}}\xspace}
\newcommand{\Oms}          {\ensuremath{\Omega^{\pm}}\xspace}
\newcommand{\degree}       {\ensuremath{^{\rm o}}\xspace}

% 5) OTHER

\newcommand{\ptjet}{\ensuremath{p_{\mathrm{T}}^{\mathrm{jet}}}\xspace}
\newcommand{\ptchjet}{\ensuremath{p_{\mathrm{T,ch\,jet}}}\xspace}
\newcommand{\ptchjetavg}{\ensuremath{\langle p_{\mathrm{T,ch\,jet}} \rangle}\xspace}
\newcommand{\RL}{\ensuremath{R_\mathrm{L}}\xspace}
\newcommand{\RLpeak}{\ensuremath{R_\mathrm{L}^\mathrm{peak}}\xspace}
\newcommand{\dif}{\ensuremath{\mathrm{d}}\xspace}
\newcommand{\corr}[1]{\ensuremath{\langle #1 \rangle}\xspace}
\newcommand{\sigeec}{\ensuremath{\Sigma_\mathrm{EEC}}\xspace}
\newcommand{\sigq}{\ensuremath{\Sigma^\mathcal{Q}_\mathrm{EEC}}\xspace}
\newcommand{\sigp}{\ensuremath{\Sigma^{++}_\mathrm{EEC}}\xspace}
\newcommand{\sigpp}{\ensuremath{\Sigma^{++}_\mathrm{EEC}}\xspace}
\newcommand{\sigm}{\ensuremath{\Sigma^{--}_\mathrm{EEC}}\xspace}
\newcommand{\sigmm}{\ensuremath{\Sigma^{--}_\mathrm{EEC}}\xspace}
\newcommand{\sigpm}{\ensuremath{\Sigma^{+-}_\mathrm{EEC}}\xspace}
\newcommand{\sigl}{\ensuremath{\sigeec^{\pm\pm}}\xspace}
\newcommand{\sigul}{\ensuremath{\sigeec^{+-}}\xspace}
\newcommand{\sigasym}{\ensuremath{\sigeec^\mathcal{Q} \, / \, \sigeec}\xspace}
\newcommand{\py}{\textsc{Pythia}\xspace}
\newcommand{\hw}{Herwig\xspace}
\newcommand{\sh}{\textsc{Sherpa}\xspace}
\newcommand{\gn}{Geant\xspace}

%%%%%%%%%%%%%%%  Title page %%%%%%%%%%%%%%%%%%%%%%%%
\begin{titlepage}
% the dates below correspond to CERN approval
% please don't touch: EB chairs will take care
\PHyear{2026}       % required, will be obtained from CERN
\PHnumber{172}      % required, will be obtained from CERN
\PHdate{10 June}  % required, will be obtained from CERN
%%%%%%%%%%%%%%%%%%%%%%%%%%%%%%%%%%%%%%%%%%%%%%%%%%%%

%%% Put your own title + short title here:
\title{Probing jet evolution with charged energy correlators in small systems}
\ShortTitle{Charged EECs in small systems with ALICE}   % appears on left page headers

%%% Do not change the next lines
\Collaboration{ALICE Collaboration\thanks{See Appendix~\ref{app:collab} for the list of collaboration members}}
\ShortAuthor{ALICE Collaboration} % appears on right page headers, do not change

\begin{abstract}

The ALICE Collaboration presents measurements of charged energy--energy correlators (charged EECs) within charged-particle jets at \fivenn in proton--proton and proton--lead collisions at the LHC. Charged EECs are a class of jet substructure observables that trace the flow of energy and electric charge within a jet, and provide a tool for disentangling the energy scales involved in the jet fragmentation process through the angular separation and charges of particle pairs. The interplay between energy distribution and charge conservation enables charged EECs to provide novel constraints on hadronization mechanisms. Measurements of charged EECs in proton--proton collisions in charged-particle jets with $20 < \ptchjet < 80$ \GeVc are compared to event generators to investigate different hadronization mechanisms and parton shower models. These model comparisons show that the like-sign EECs are sensitive to changes in parton shower modeling, while unlike-sign EECs are sensitive to different hadronization schemes. Measurements in proton--lead collisions indicate that cold nuclear matter effects on charged EECs are consistent with charge-independent behavior.

\end{abstract}
\end{titlepage}

\setcounter{page}{2} %please do not remove this line

%%%%%%%%%%%%%%%%%%%%%%%%%%%%%%%%
% begin main text
%%%%%%%%%%%%%%%%%%%%%%%%%%%%%%%%

\section{Introduction}

Jets of hadrons are produced in particle collisions at high-energy colliders, and their evolution spans a wide range of energy scales within quantum chromodynamics (QCD). They arise from highly energetic quarks and gluons (collectively termed partons), produced in hard-scattering processes between the constituents of the colliding particles. These partons undergo successive QCD splittings, developing into a parton shower through gluon radiation and gluon splitting into quark–antiquark pairs. When the resulting partons reach a characteristic energy scale $\Lambda_\mathrm{QCD}$~\cite{lambda}, they combine into color-neutral hadrons in a process known as hadronization. The resulting final-state hadrons can then be clustered into jets with jet clustering algorithms~\cite{antikt}. The evolution of a jet thus involves multiple QCD regimes through various energy scales, from the large momentum transfer of the initial scattering (at up to hundreds of \GeV) to hadronization (at around $\Lambda_\mathrm{QCD}$). Therefore, an accurate description of the internal structure of QCD jets demands a full understanding of QCD itself at all scales. Disentangling the contributions from dynamics at different scales requires observables that are sensitive to different stages of the jet evolution across this hierarchy of scales.

The two-point energy--energy correlator (EEC) is an observable that tracks the energy flow in a jet. Originally proposed as a basis for precision studies of QCD~\cite{eecee}, EECs have recently received renewed theoretical and experimental interest, thanks to the large jet production cross sections at hadron colliders combined with high-luminosity datasets, enabling precise measurements of EECs in the collinear limit (i.e. at small angles). EECs have been measured in a variety of contexts: at the Relativistic Heavy Ion Collider (RHIC)~\cite{star} and the Large Hadron Collider (LHC), from small~\cite{star,aliceeecpp,cmspp,aliceppb} to large collision systems~\cite{cmspbpb}, and in inclusive and D$^0$ meson-tagged jets~\cite{aliceeecd0}. Applications of EECs at hadron colliders include precision extractions of the strong coupling constant~\cite{cmspp,atlasteec}, gluon polarimetry~\cite{spin}, and modifications of jet substructure in quark--gluon plasma, at both early~\cite{early} and late times~\cite{ceec_ratio_th} in the fireball evolution. For a recent comprehensive review, see~\cite{eec-rev}.

The EEC (denoted as \sigeec in this manuscript) is defined as an energy-weighted cross section of charged-particle pairs in jets, measured as a function of the pair opening angle $\RL = \sqrt{\Delta\varphi^2 + \Delta\eta^2}$, where $\Delta\varphi$ and $\Delta\eta$ are the differences in azimuthal angle and pseudorapidity, respectively, between the two particles. The experimental definition used in this measurement is
\begin{equation}
\label{eq:eec}
    \sigeec(\RL) = \frac{1}{ N_{\mathrm{jet}}\cdot\Delta}\int_{\RL-\frac{1}{2}\Delta}^{\RL+\frac{1}{2}\Delta}\dif\RL' \,\, \sum_{\mathrm{jets}} \,\,\, \sum_{i, j \in \mathrm{jet}} \frac{p_{\mathrm{T},i}}{\ptjet}\frac{p_{\mathrm{T},j}}{\ptjet} \delta(\RL'-R_{\mathrm{L}, ij}),
\end{equation}
where $N_\mathrm{jet}$ is the number of jets, $\Delta$ is the angular bin width, the indices $i$ and $j$ denote the jet constituents and run over all constituents\footnote{Note that the pairs $(i,j)$ and $(j,i)$ are considered to be distinct; and that self-correlations, i.e. $(i,i)$, have $\RL=0$ and are thus outside of the reported \RL range.}, $p_{\mathrm{T},i}$ and $p_{\mathrm{T},j}$ are the transverse momenta of particles $i$ and $j$ respectively, $\ptjet$ is the jet transverse momentum, $\delta$ is the Dirac delta function, and $R_{\mathrm{L},ij}$ is the angular distance between particles $i$ and $j$. This measurement uses jets built from charged particles only, whose energies are not directly accessible with the ALICE detector. Rather than approximating it by assuming a pion mass, as was done in other experimental measurements~\cite{star}, we redefine the observable using the directly measurable transverse momentum in place of energy.

As a jet-substructure observable, the EEC is useful to potentially separate the various energy scales of QCD. This can be seen as a consequence of the angular ordering in QCD parton emissions in a parton shower, where wider-angle emissions are associated with larger energy scales, where $\alpha_\mathrm{S}$ is small and perturbative calculations are reliable. Consequently, EECs can be considered as having three angular regions. At large opening angle, the dynamics of strongly-interacting partons at the higher energy scales of QCD result in strong correlations between energy and angle, and perturbative QCD (pQCD) calculations predict a scaling as $\sigeec \sim 1/\RL$~\cite{ceec_th}. At small opening angle, the EEC is dominated by hadron-level correlations and is expected to scale linearly with angle: $\sigeec \sim \RL$~\cite{eecth}. In between, there is a transition region and a peak in the EEC, which characterizes the hadronization transition from the partonic to hadronic degrees of freedom.

The EEC takes advantage of how QCD induces correlations between the angles and energies of the different particles within a jet, which likely have a common origin from the initial hard-scattered parton. However, any conserved quantity is also correlated with these properties. The charged EECs, proposed in~\cite{ceec_th}, incorporate the correlations of electric charge with energy and angle in two ways. First, instead of considering all charged-particle pairs, the charge-selected EECs are calculated with \cref{eq:eec} but only with particle pairs that satisfy the specific charge combination of interest. For example, the unlike-sign EEC (denoted as \sigpm in this manuscript) only considers pairs where one of the particles has positive charge, and the other has negative charge. The two like-sign EECs (denoted as \sigp and \sigm in this manuscript) are defined likewise. Second, the charge-weighted EEC (denoted as \sigq in this manuscript) incorporates the product of the particle charges into the energy weighting, and is defined as
\begin{equation}
\label{eq:eecq}
    \sigq(\RL) = \frac{1}{ N_{\mathrm{jet}}\cdot\Delta}\int_{\RL-\frac{1}{2}\Delta}^{\RL+\frac{1}{2}\Delta}\dif\RL' \,\, \sum_{\mathrm{jets}} \,\,\, \sum_{i, j \in \mathrm{jet}} q_i q_j \frac{p_{\mathrm{T},i}}{\ptjet}\frac{p_{\mathrm{T},j}}{\ptjet} \delta(\RL'-R_{\mathrm{L}, ij}),
\end{equation}
where $q_i$ and $q_j$ are the charges of particles $i$ and $j$, respectively. Thus, like-sign pairs contribute positive energy weights (since $q_iq_j > 0$) and unlike-sign pairs contribute negative energy weights (since $q_iq_j < 0$). The charge-weighted EEC can thus be expressed in terms of the charge-selected EECs as\footnote{\Cref{eq:eecq-sel} is only exact if all charged particles in the jet are singly charged. However, as multi-charged hadrons in jets are very rare, this expression is valid to a very good approximation.}
\begin{equation}
\label{eq:eecq-sel}
    \sigq = \sigp + \sigm - \sigpm.
\end{equation}
To refer collectively to the charge-selected and charge-weighted EECs, the term ``charged EECs'' will be used. To distinguish \sigeec, as defined in \cref{eq:eec}, from the charged EECs, it will be referred to as the inclusive EEC in this manuscript. Like the charge-weighted EEC, the inclusive EEC can also be expressed in terms of the charge-selected EECs,
\begin{equation}
\label{eq:eec-sel}
    \sigeec = \sigp + \sigm + \sigpm,
\end{equation}
since the three subsets relevant for each charge-selected EEC (unlike-sign, both positive charge, both negative charge) are disjoint and their union consists of all possible pairs.

From the relations in \cref{eq:eec-sel,eq:eecq-sel}, the relative contribution of unlike-sign and like-sign pairs to the inclusive EEC can be quantified by computing the ratio of the charge-weighted EEC to the inclusive EEC, the charged EEC asymmetry \sigasym:
\begin{equation}
\label{eq:eec-asy}
    \frac{\sigq}{\sigeec} = \frac{\sigp + \sigm - \sigpm}{\sigp + \sigm + \sigpm} = \frac{\sigp + \sigm}{\sigp + \sigm + \sigpm} - \frac{\sigpm}{\sigp + \sigm + \sigpm}.
\end{equation}
This charged EEC asymmetry is the ratio of the difference of the like-sign and unlike-sign EECs to their sum. It quantifies the relative contribution of unlike-sign and like-sign pairs to the overall inclusive EEC. For example, a charged EEC asymmetry value of $-0.2$ means that the relative contributions of the like-sign and unlike-sign EEC to the inclusive EEC are 40\% and 60\%, respectively. One can also test the charge dependence of jet fragmentation via the like-sign EEC ratio $\sigp \, / \, \sigm$. For example, a ratio of unity would indicate that jets fragment equally into positively-charged and negatively-charged particles.

Ratios of charged EECs can also benefit from large cancellations in systematic uncertainties, yielding more precise measurements. Charged EECs may offer stronger sensitivity than the inclusive EEC to non-perturbative phenomena like hadronization, as they track not only the flow of energy but also the flow of charge within a jet~\cite{ceec_th}. As a result, this new observable may be a useful probe of both the perturbative and non-perturbative aspects of jets.

The hard scattering cross section and the first few emissions in the parton shower are calculable with pQCD techniques. However, an exact description of low-energy phenomena, in particular hadronization, from first-principles QCD remains elusive. Monte Carlo (MC) event generators that simulate particle collisions thus often rely on QCD-inspired phenomenological models to produce a set of colorless hadrons from the color-charged partons that arise from a parton shower. There are two general classes of hadronization models: those based on the idea of breaking one-dimensional QCD flux tubes, termed Lund string-breaking models~\cite{lund1,lund2}; and those based on treating partons close in phase space as clusters that fragment into hadrons, termed cluster models. The development of Lund string models has progressed primarily in the context of the \py event generator~\cite{py83}, while two different implementations of cluster hadronization have evolved within \hw~\cite{herwig73} and \sh~\cite{sherpa3}. Comparisons against a variety of measurements therefore serve several purposes. First, they can corroborate and constrain new methods and techniques in shower modeling. Second, they can discriminate among hadronization models. Third, they can establish how sensitive an observable is to different ways that the evolution of a jet can be modeled. The comparison of charged EEC measurements with different event generators gives new insights into non-perturbative QCD phenomena.

In addition, the differential nature of the charged EECs can help investigate modifications in different collision systems. Previous ALICE measurements of the inclusive EEC~\cite{aliceppb} show that the EEC is modified in \pPb collisions relative to \pp collisions, with a suppression of the EEC distribution at small angles and an enhancement at large angle. The charge dependence of this modification can be tested with charged EECs.

This paper presents the first LHC measurements of energy--energy correlators for charged particles in charged-particle jets, using their opening angle and their charges to separate the energy scales of the jet evolution. This analysis utilizes the charged-particle tracking systems of the ALICE experiment to measure the inclusive EEC, the charged EECs, the charged EEC asymmetry \sigasym, and the like-sign EEC ratio $\sigp \, / \, \sigm$, as a function of the pair opening angle \RL in charged-particle jets with $20 < \ptchjet < 80$ \GeVc in proton--proton (\pp) collisions at \five. In addition, the inclusive EEC and charged EECs are also measured in proton--lead (\pPb) collisions at \fivenn. These measurements are compared to a corresponding \pp baseline measurement, distinct from the \pp measurement mentioned above, which is treated with the same background subtraction procedure as in \pPb collisions.

\Cref{sec:data} describes the experimental setup and the data used in the analysis. \Cref{sec:methods} describes the analysis method, including the detector effect corrections and background subtraction, and \Cref{sec:syst} addresses the associated systematic uncertainties. \Cref{sec:results} compares the measurements to predictions from Monte Carlo event generators, as well as state-of-the-art pQCD calculations. Detailed studies of the transition region are also presented.
\section{The ALICE detector and data}
\label{sec:data}

This analysis makes use of data collected in \pp and \pPb collisions during LHC Run 2 at \fivenn in 2017 and 2016, respectively, the same data as those in previous ALICE EEC measurements in \pp~\cite{aliceeecpp} and \pPb collisions~\cite{aliceppb}. Minimum-bias events are identified by coincident signals in the V0A and V0C scintillator hodoscopes~\cite{alice-v0}, located at far forward ($2.8 < \eta < 5.1$) and far backward ($-3.7 < \eta < -1.7$) pseudorapidity, respectively. To ensure optimal and uniform detector performance, reconstructed primary vertices are required to be within 10 cm of the nominal interaction point along the beam direction. Further selections are performed on event properties to minimize pileup~\cite{aliceeecpp}.  Charged-particle tracks with $0.15 < \pt < 100$ \GeVc and $|\eta| < 0.9$ are reconstructed with the central barrel tracking systems: the Inner Tracking System (ITS)~\cite{aliceits} and the Time Projection Chamber (TPC)~\cite{alicetpc}. Event and track selections are identical to those in~\cite{aliceeecpp}. The relative \pt resolution is approximately 1\% for tracks with $\pt \approx 1$ \GeVc and rises to 4\% for $\pt > 50$ \GeVc~\cite{aliceeecd0}.  The angular resolution is around 1 mrad for tracks with $\pt = 1$ \GeVc and improves to less than 0.6 mrad for tracks with $\pt > 5$ \GeVc~\cite{aliceeecpp}. A detailed description of the ALICE detector and its general performance in LHC Run 2 can be found in~\cite{aliceperf}. 
\section{Experimental method}
\label{sec:methods}

Reconstructed charged-particle tracks are clustered into anti-$k_\mathrm{T}$ jets~\cite{antikt} with jet radius parameter $R=0.4$ and $E$-scheme recombination~\cite{escheme} with the FastJet package (version 3.3.4)~\cite{fj}. For the purposes of jet finding, all particles are assumed to have the charged-pion mass. Only jets fully within the TPC acceptance of \etarange{0.9} are considered in this analysis, i.e. $|\eta_\mathrm{jet}| < 0.9 - R = 0.5$.  For the purposes of pairing particles for the EEC, only jet constituents with $\pt > 1$ \GeVc are considered, to efficiently eliminate contributions from pairs in which one or both particles are from the underlying event (UE). This value of the cutoff is commonly used in EEC measurements~\cite{aliceeecpp,cmspp,aliceeecd0,aliceppb,cmspbpb}. In \pp collisions, no other steps are taken to account for the UE.  Results from \pPb data, as well as the \pp baseline measurement used to compare to the \pPb data, incorporate two additional background subtraction methods applied to each of the charged and inclusive EECs: an area-based $\ptchjet$ correction described in~\cite{rho,jetarea}, as well as a perpendicular cone subtraction that removes contributions from pairs that include a particle from the UE. This perpendicular cone subtraction method is applied as introduced in~\cite{aliceppb}: two cones with radius of $R = 0.4$ that are perpendicular in azimuth to the jet axis are constructed, and all particles within the cones are assumed to originate from the UE. The contributions in the jet arising from pairs with one hard-scattering track and one UE track (signal--background) and from pairs with two UE tracks (background--background) can then be estimated by creating pairs between jet particles and cone particles, particles within the same cone, and between particles in different cones. This is done by azimuthally rotating each perpendicular cone's axis such that it coincides with the jet axis. Three sets of EECs are then calculated from particular kinds of pairs, using the area-subtracted \ptchjet of the original jet in the denominator of the pair's energy weight. The jet-cone EEC is calculated only from pairs where one particle is from a cone and the other from the jet, averaged over the two cones,
\begin{equation}
\sigeec^\text{jet-cone} = \frac{1}{2} (j \times c_1 + c_1  \times j + j \times c_2 + c_2  \times j ),
\end{equation}
where $j$ represents a track from the jet, $c_1$ and $c_2$ represent a track from either perpendicular cone, and $\times$ represents constructing an EEC pair. The cone-cone EEC is calculated only from pairs where both particles are from the same cone, also averaged over the two cones:
\begin{equation}
\sigeec^\text{cone-cone} = \frac{1}{2}(c_1 \times c_1 + c_2 \times c_2).
\end{equation}
The mixed cone EEC is calculated only from pairs where the particles are from different cones:
\begin{equation}
\sigeec^\text{mixed cone} = c_1 \times c_2 + c_2 \times c_1.
\end{equation}
The perpendicular cone-subtracted EEC is then estimated as
\begin{equation}
\sigeec^\mathrm{sub} = \sigeec - \sigeec^\text{jet-cone} - \sigeec^\text{cone-cone} + \sigeec^\text{mixed cone}.
\end{equation}

The jet-cone EEC removes the signal-background correlations, and the cone-cone EEC removes the background-background correlations in the raw EEC. However, the jet-cone EEC not only contains signal-background correlations, but also background-background correlations, which should not be subtracted twice. These background-background correlations are added back in via the mixed cone EEC.

Detector effects are estimated with simulated collision events that are propagated through a GEANT3~\cite{geant} representation of the ALICE detector reproducing its response. Hard scattering events with \textsc{Pythia} 8.243~\cite{py82} (Monash 2013 tune~\cite{monash}) and $\hat{p}_\mathrm{T}$ (the transverse momentum of the hard scattering) greater than 5 \GeVc were used for \pp collisions, and minimum-bias collisions with DPMJET (version 3.0-5)~\cite{dpmjet1,dpmjet2,dpmjet3} were used for \pPb collisions. The reconstruction of the simulated collision events is anchored to the detector conditions during the relevant data-taking periods.

Because the ALICE track-pair angular resolution is much smaller than the width of the $\RL$ intervals, no angular bin migrations are considered in the corrections~\cite{aliceeecpp}. A two-dimensional bin-by-bin correction is employed to account for migrations in $\ptchjet$, the pair energy weights, and detector efficiency. This is the same method used in previous ALICE EEC measurements~\cite{aliceeecpp,aliceeecd0,aliceppb}, and optimizes the statistical precision of the corrections~\cite{aliceeecpp}.

The detector effects that affect charged EECs are the reconstructed track \pt resolution and the reconstruction efficiencies for single tracks and track pairs.  The track \pt resolution alters the energy weight for reconstructed pairs.  The single-track reconstruction efficiency skews the \ptchjet of a reconstructed jet to lower values on average relative to the true \ptchjet, which also alters the energy weight, and reduces the number of pairs in the EECs when one or both tracks in the pair are not reconstructed. The efficiency of track-pair reconstruction can also decrease when the two particles have the same charge sign and similar \pt, pseudorapidity, and azimuthal angle. Since they trace similar trajectories within the TPC volume, the partially- or fully-overlapping clusters in the TPC can cause the two trajectories to be reconstructed as a single track, or as no track at all. Previous ALICE analyses in EECs~\cite{aliceeecpp} and correlation femtoscopy~\cite{femto} have shown that the pair efficiency is largely driven by the single-track reconstruction efficiency at large angular separation, but drops dramatically for pairs with $\RL < 0.01$ (with decreases of up to 30\% for pairs at higher \pt). Consequently, the charged EECs are measured as a function of the pair opening angle \RL from 0.01 to 0.4 and are reported in three intervals of $\ptchjet$: 20--40 \GeVc, 40--60 \GeVc, and 60--80 \GeVc.

Another possible detector effect is the charge-reconstruction inefficiency (the likelihood for a track to be reconstructed with the wrong charge sign), which can shift track pairs from one charge-selected EEC to another. The inefficiency was estimated in Monte Carlo simulation by measuring the frequency of charge misreconstruction for reconstructed particles, i.e. the particle was reconstructed but with the opposite charge sign. The charge-reconstruction inefficiency for single tracks was found to be fully consistent with zero within uncertainties.
\section{Systematic uncertainties}
\label{sec:syst}

The main sources of systematic uncertainty arising from detector effects are the determination of the reconstruction efficiency of single tracks and of track pairs. 

Two sources of systematic uncertainty related to the bin-by-bin correction were evaluated: the dependence on the choice of Monte Carlo generator used to derive the bin-by-bin correction, and the use of a Bayesian unfolding technique instead of a bin-by-bin correction. The sources of systematic uncertainties, and the methods used to estimate them, are identical to those in previous ALICE EEC measurements in \pp~\cite{aliceeecpp} and \pPb collisions~\cite{aliceppb}.

By varying track-selection criteria, previous ALICE studies have determined that the uncertainty in the single-track reconstruction efficiency is approximately 3\%~\cite{tracking}. Following the methodology established in~\cite{aliceeecpp}, 3\% of reconstructed tracks are randomly removed and the bin-by-bin correction is recalculated. The difference with respect to the nominal bin-by-bin correction is used as the systematic uncertainty.

The pair reconstruction efficiency systematic uncertainty was determined by imposing various pair-selection parameters on the reconstructed track pairs and studying the resulting variation in the EECs.

The dependence on the event generator used to derive the bin-by-bin correction was evaluated by comparing correction factors derived from \py 8.235 and \hw 7.2.1 coupled with the same parametrized detector response.

As an alternative to bin-by-bin correction, the EECs were also unfolded using an iterative Bayesian method~\cite{bayesian}. The difference between the two approaches was taken as a systematic uncertainty. The EECs were unfolded with a Bayesian unfolding technique implemented in the RooUnfold package (version 3.0.5)~\cite{roounfold} and compared to the bin-by-bin corrected result to determine the unfolding uncertainty.  The systematic uncertainties for the measurement in \pp collisions for 20--40 \GeVc jets are shown in \Cref{tab:syst-pp}, and are similar in higher \ptchjet intervals. A summary of the systematic uncertainties for the measurement in \pPb collisions is shown in \Cref{tab:syst-ppb}.

\begin{table}[!ht]
\caption{A summary of the relative systematic uncertainties for the inclusive, charge-selected, and charge-weighted EECs, as well as the charged EEC asymmetry and the like-sign EEC ratio in \pp collisions for 20--40 \GeVc charged-particle jets.}
\centering
\begin{tabular}{l|ccccccc}
                     & \sigeec & \sigp  & \sigm  & \sigpm & \sigq  & \sigasym & $\sigp \, / \, \sigm$ \\ \hline
Tracking efficiency  & 0--2\%  & 0--2\% & 0--2\% & 0--2\% & 0--4\% & 0--3\%   & 0--1\%                \\
Pair efficiency      & 0--1\%  & 0--1\% & 0--1\% & 0--1\% & 0--1\% & 0--1\%   & 0--1\%                \\
Generator dependence & 1--2\%  & 1--3\% & 1--3\% & 2\%    & 2--3\% & 1--2\%   & 1--2\%                \\
Unfolding            & 0--5\%  & 1--3\% & 1--5\% & 1--4\% & 0--3\% & 1--11\%  & 0--1\%                \\ \hline
Total                & 2--5\%  & 2--5\% & 2--6\% & 2--5\% & 3--6\% & 2--11\%  & 1--3\%
\end{tabular}
\label{tab:syst-pp}
\end{table}

\begin{table}[!ht]
\caption{A summary of the relative systematic uncertainties for the charged EECs in \pPb collisions and for the ratio of charged EECs in \pPb to \pp collisions.}
\centering
\begin{tabular}{l|ccccccc}
              & \sigp  & \sigm   & \sigpm  & \sigq  \\ \hline
20--40 \GeVc  & 3--6\% & 2--9\%  & 2--8\%  & 3--13\% \\
40--60 \GeVc  & 4--8\% & 2--11\% & 2--8\%  & 4--15\% \\
60--80 \GeVc  & 5--8\% & 3--10\% & 3--10\% & 7--19\% \\
Ratio         & 4--6\% & 3--10\% & 3--9\%  & ---     \\
\end{tabular}
\label{tab:syst-ppb}
\end{table}

For the inclusive EEC, the charged EECs, the charged EEC asymmetry, and the like-sign EEC ratio in \pp collisions, the relative difference between the baseline measurement and the systematic variation was taken as the systematic uncertainty for each source. The total systematic uncertainty was calculated as the sum in quadrature of all sources considered, and is below 6\% for the inclusive and charge-selected EECs in the three \ptchjet intervals, below 12\% for the charge-weighted EEC, below 11\% for the charged EEC asymmetry, and below 8\% for the like-sign EEC ratio. The uncertainties increase slightly for larger \ptchjet.

For the \pPb measurement and its corresponding \pp baseline, the same methodology utilized in~\cite{aliceppb} was used to estimate the single-track efficiency systematic uncertainty. It was found to be dependent on \pt: 1--2\% for particles with $\pt < 5$ \GeVc and rising up to 2.5\% for particles with $\pt > 8$ \GeVc. To estimate the generator dependence of the EEC correction factor, the nominal bin-by-bin correction extracted from the DPMJET simulation was compared to \pPb \py 6.425 simulations~\cite{py64} (with the Perugia 2011 tune~\cite{perugia}) paired with an underlying event simulated with HIJING~\cite{hijing} and a GEANT3 detector simulation~\cite{geant}. The relative difference between the bin-by-bin corrections derived from the two simulations were taken as the uncertainty, and are less than 12\% for the charge-selected EECs in \pPb collisions. The uncertainties relating to the unfolding and pair reconstruction efficiency were estimated with procedures identical to those described above for the \pp measurement.

In addition, three uncertainties due to the background subtraction were included for both the \pPb measurement, its corresponding \pp baseline, and the EEC ratio in \pPb to \pp. The uncertainty arising from the choice of perpendicular cone size was estimated as follows: the area of the perpendicular cone was matched to the area of the jet (rather than using a fixed circular cone of radius 0.4), estimated with the Voronoi method in the FastJet package~\cite{fj}, and compared to the default method. This systematic uncertainty is less than 5\% in \pPb collisions, and less than 2\% in the EEC ratio. To estimate the effect of background fluctuations within the perpendicular cones, the jet-cone and cone-cone distributions were estimated with only one of the cones, and the relative difference between using only one cone and using only the other was used as the systematic uncertainty, resulting in an uncertainty of less than 7\% in \pPb collisions, less than 3\% in the \pp baseline measurement, and less than 2\% in the EEC ratio. To estimate the effect of background sparsity on the determination of the background density $\rho$ in the area-based \ptchjet subtraction in \pp collisions, the \ptchjet was reduced by the average $\rho A_\mathrm{jet}$ in \py simulations and compared to the unsubtracted EEC. This systematic uncertainty was determined to be 3\%, independent of \RL. For each of the \pPb and \pp baseline EECs, the different sources of uncertainty were assumed to be uncorrelated and added in quadrature to obtain the total uncertainty. The pair efficiency, unfolding, and all three background subtraction systematic uncertainties were assumed to be correlated between the \pp and \pPb data sets, and taken into account as such in the ratio. The single-track efficiency and generator-dependence systematics were assumed to be uncorrelated between \pPb and \pp collisions and added in quadrature for the ratio. The total systematic uncertainty for the charged and inclusive EECs in \pPb collisions is generally less than 13\%, and is less than 10\% for the charged and inclusive EEC ratio in \pPb to \pp collisions.
\section{Results and discussion}
\label{sec:results}

\begin{figure}[!htb]
\centering
\includegraphics[width=0.99\textwidth]{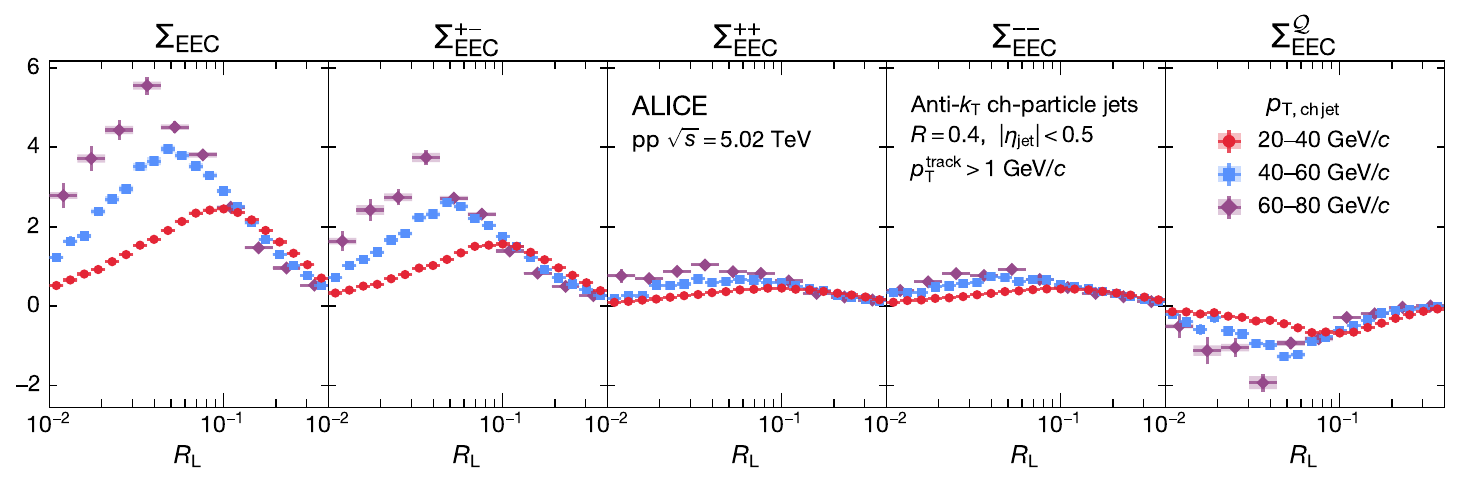}
\caption{The inclusive (left panel), charge-selected (middle three panels), and charge-weighted EECs (right panel) in \pp collisions are plotted as a function of \RL for 20--40 (red circles), 40--60 (blue squares), and 60--80 (purple diamonds) \GeVc charged-particle jets. Statistical uncertainties are shown as vertical lines, and systematic uncertainties as shaded boxes. }
\label{fig:eec-pp}
\end{figure}%

\begin{figure}[!htb]
\centering
\includegraphics[width=0.99\textwidth]{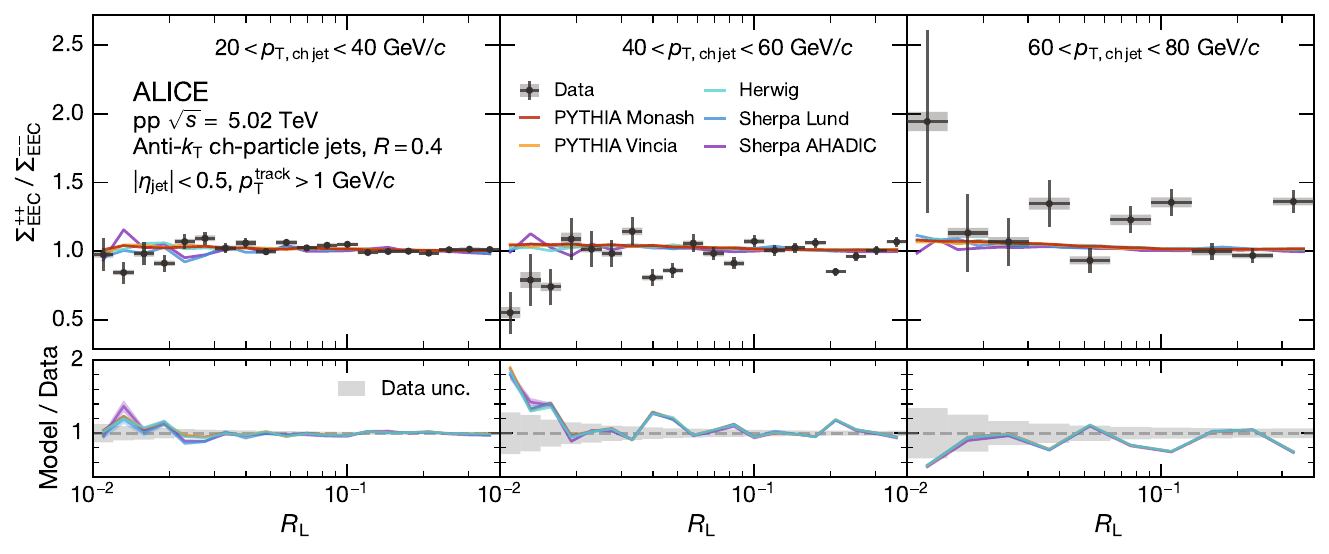}
\caption{The ratio of \sigpp to \sigmm in \pp collisions as a function of \RL for 20--40, 40--60, and 60--80 \GeVc charged-particle jets, along with predictions from five MC event generators. Statistical uncertainties on the data are shown as vertical lines, and systematic uncertainties as shaded boxes. Ratios to the data are shown in the second row, and the relative combined statistical and systematic uncertainties on the data are shown as shaded gray boxes surrounding the dashed unity line.}
\label{fig:eec-lratio}
\end{figure}

The inclusive and charged EECs in \pp collisions, measured as a function of the pair angular separation \RL, is shown in three \ptchjet intervals of 20--40 \GeVc, 40--60 \GeVc, and 60--80 \GeVc in \Cref{fig:eec-pp}. Each panel corresponds to the inclusive (left), charge-selected (middle three), and charge-weighted (right) EEC, and includes the three \ptchjet intervals. The charge-selected EECs, from the second to fourth panels of \cref{fig:eec-pp}, show the same distinct angular domains as the inclusive EEC: a hadronic or non-perturbative region at small angles (where the distributions increase), a partonic or perturbative region at large angles (where the distributions decrease), and a hadronization transition region at intermediate angles (near the peak of the distributions). As in the inclusive EEC, the angle corresponding to the peaks of the charge-selected EECs decrease as \ptchjet increases. The overall negative sign of the charge-weighted EEC indicates that unlike-sign pairs dominate the contributions to the inclusive EEC compared to like-sign pairs, according to \cref{eq:eecq-sel}.  This can be understood purely from estimating the total number of like-sign and unlike-sign pairs within a jet. At LHC energies, the majority of inclusive jets are initiated by gluons, which have zero electric charge. These jets are likely to have the same number of positively-charged and negatively-charged particles. Assuming a jet with $n$ positively-charged and $n$ negatively-charged particles, the number of unlike-sign pairs with nonzero opening angle is $2n^2$, but the number of like-sign pairs with nonzero opening angle is fewer, at $2n(n-1)$. The dominance of gluon-initiated jets in inclusive jet samples also results in a like-sign EEC ratio near unity, shown in \cref{fig:eec-lratio}. The measured like-sign EEC ratios are also consistent with five MC models, which are shown as colored lines and described in more detail in \cref{sec:comp-mc}.

\subsection{Comparison to Monte Carlo event generators}
\label{sec:comp-mc}

\begin{figure}[!htb]
\centering
\includegraphics[width=0.99\textwidth]{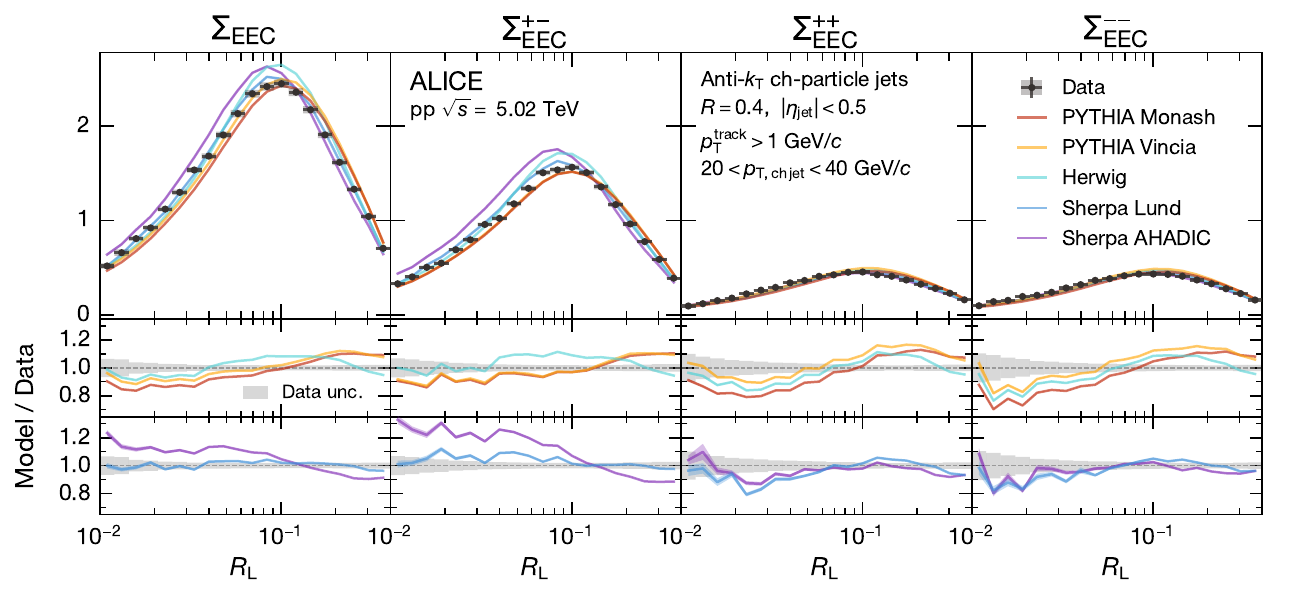}
\caption{Comparison of the inclusive and charge-selected EECs in \pp collisions for 20--40 \GeVc jets with five different MC models. In the first row, statistical uncertainties on the data are shown as vertical lines, and systematic uncertainties are shown as shaded boxes. Statistical uncertainties on the five models are shown as shaded bands around the lines but are mostly too small to be seen. For the model-to-data ratios displayed in the bottom two rows, the relative combined statistical and systematic uncertainties on the data are shown as shaded gray boxes surrounding the dashed unity line. The relative statistical uncertainties from the models are shown as colored shaded bands around each line.}
\label{fig:eec-mc}
\end{figure}%

The sensitivity of a particular observable to the modeling of jet evolution mechanisms can be tested by comparing predictions from Monte Carlo generators. \Cref{fig:eec-mc} compares the experimental measurements of the charged and inclusive EEC to predictions from five Monte Carlo event generators, using the same jet-algorithm parameters and \pt cutoff as the data. Only the lowest \ptchjet interval, 20--40 \GeVc, is compared to MC, as uncertainties at higher \ptchjet become too large to distinguish between models. In the second row of \cref{fig:eec-mc}, \py (version 8.315~\cite{py83} with the Monash 2013 tune~\cite{monash}, in red), which implements string breaking; and \hw (version 7.3.0 with the default tune~\cite{herwig++,herwig73}, in light blue), which implements cluster hadronization, are compared to the data. For the inclusive EECs, the two models differ: \py and \hw have a very similar shape for $\RL \lesssim 0.04$, but diverge at larger angles. The shape of the ratio for inclusive EECs closely follows that of the unlike-sign pairs. Consequently, the disagreement in the inclusive EECs appears to be due to the unlike-sign pairs, as the shape of the model-to-data ratios for \py and \hw are similar for the like-sign EECs.

Not only do \py and \hw model hadronization differently, but they also take different approaches to the parton shower: \py takes a \pt-ordering approach, while \hw 7.3 implements angular ordering. In order to separate the effects of jet evolution in the perturbative regime from hadronization without convolving effects from both, two studies are presented in the second and third rows of \cref{fig:eec-mc}. In the second row, the hadronization model is kept the same and two independent parton showers are employed, and in the third row, the parton shower model is kept the same and two independent hadronization schemes are employed. These studies demonstrate the sensitivity of the charged EECs to the modeling of each stage of a jet.

The second row of \cref{fig:eec-mc} compares two parton shower models implemented within \py: the default Simple shower~\cite{simple} with the Monash 2013 tune~\cite{monash} (in red) and the Vincia antenna shower~\cite{vincia1,vincia2,vincia3,vinciaqed} with the dedicated Vincia tune (in orange). The default Lund string hadronization model implemented in \py is maintained. The two models yield near-identical unlike-sign EECs, and although the shapes of the like-sign EECs are very similar, they differ numerically by up to 10\%. In addition, the negative sign of the charge-weighted EECs indicate that unlike-sign pairs contribute more to the overall inclusive EEC than like-sign pairs do. Because only the like-sign correlations show sensitivity to the parton shower model, the inclusive EEC is less sensitive to the parton shower model compared to the like-sign EEC. Some of the like-sign sensitivity to the shower model observed in these comparisons may be due to the differences between the Monash and Vincia tunes, rather than the model itself. However, it should be noted that the Vincia tune is not a full \py tune, but rather a tune of a subset of hadronization and multi-parton interaction parameters only.

Hadronization effects on the EEC are isolated with \sh (version 3.0.3~\cite{sherpa3}). Using the default \sh shower model~\cite{css}, two hadronization schemes are compared in the third row of \cref{fig:eec-mc}. 
The native hadronization model in \sh, AHADIC++~\cite{ahadic} (in purple), hadronizes final-state partons with a cluster hadronization algorithm somewhat similar to that used in \hw. \sh also provides a direct interface to \py 8's string-breaking scheme\footnote{In this analysis, \sh is interfaced with \py 8.315.}, which is termed the Lund model\footnote{We note that \sh itself uses the name ``Pythia'' to refer to this setting, but to avoid confusion with the comparisons to pure \py elsewhere in this text, this text will instead use the term ``Lund'' to refer to this particular \sh configuration.} in this text (shown in dark blue). In both cases, the default \sh tune is used and all \sh parameters are kept the same, with only the hadronization models swapped. A reversed dependence compared to the \py comparisons is observed: only the unlike-sign EEC shows sensitivity to the specific hadronization model; the like-sign EEC remains relatively unchanged between \sh AHADIC and Lund.

Taken together, the \sh and \py comparisons indicate that differences in shower and hadronization modeling seem to only modify specific charge-selected EECs: from \sh, differences in hadronization manifest predominantly in the unlike-sign EEC, and not in the like-sign; and from \py, differences in the parton shower manifest in the like-sign EEC, and not in the unlike-sign. Because of the dominance of the unlike-sign EEC over the like-sign (demonstrated by the negative charge-weighted EEC), the inclusive EEC tends to be less sensitive overall to the parton shower than the like-sign EEC. One might expect that unlike-sign EECs are sensitive to the hadronization mechanism because string breaking conserves charge at each break, while cluster hadronization conserves charge in the aggregate. On the other hand, it is unclear why like-sign correlations seem immune to these differences. In addition, changes in how the parton shower is modeled, in principle, should change the characteristics of the final-state partons resulting from the shower, which should propagate in some way to the final-state hadrons. The observation that only like-sign pair correlations seem to be modified, rather than all charge-selected EECs, is not yet understood.

\begin{figure}[!htb]
\centering
\includegraphics[width=0.99\textwidth]{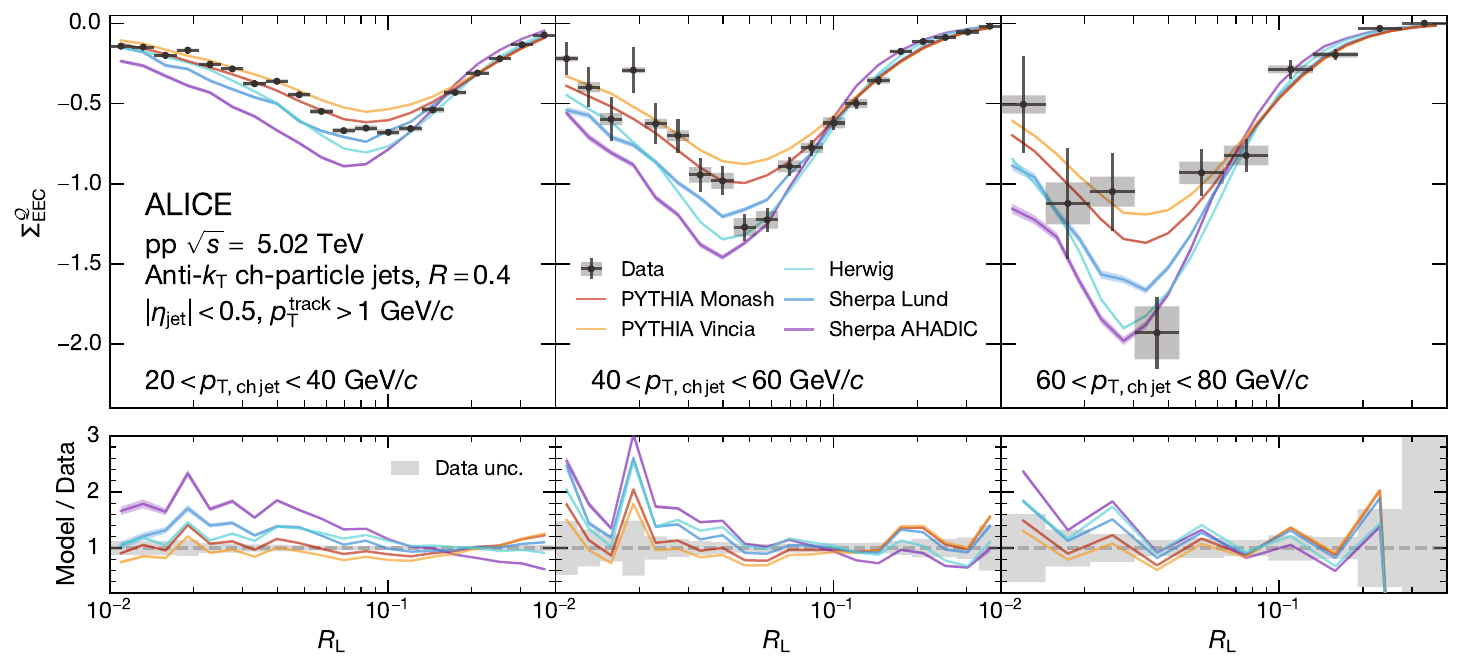}
\caption{Comparison of the charge-weighted EEC in \pp collisions with five different Monte Carlo models in each interval of \ptchjet. In the first row, statistical uncertainties on the data are shown as vertical lines, and systematic uncertainties are shown as shaded boxes. Statistical uncertainties on the five models are shown as shaded bands around the lines but are mostly too small to be seen. For the model-to-data ratios displayed in the bottom two rows, the relative combined statistical and systematic uncertainties on the data are shown as shaded gray boxes surrounding the dashed unity line. The relative statistical uncertainties from the models are shown as colored shaded bands around each line.}
\label{fig:eecQ-mc}
\end{figure}%

When the charge-selected EECs are combined into the charge-weighted EEC, model differences tend to be magnified, as shown in \cref{fig:eecQ-mc}. Models differ from the data up to 200\%. In particular, only the \py models are properly able to capture the overall balance between unlike-sign and like-sign pairs. Both \sh models, and AHADIC especially, overpredict the prevalence of unlike-sign pairs, especially at low \ptchjet.

Overall, \sh Lund and \hw best describe the charge-selected and inclusive EECs in the data, to within 10\%. In contrast, the \py models are better descriptions of the charge-weighted EEC. Ultimately, no model provides a consistent description across all observables.

\subsection{Characterization of the transition region in \pp collisions}

\begin{figure}[!htb]
\centering
\includegraphics[width=0.99\textwidth]{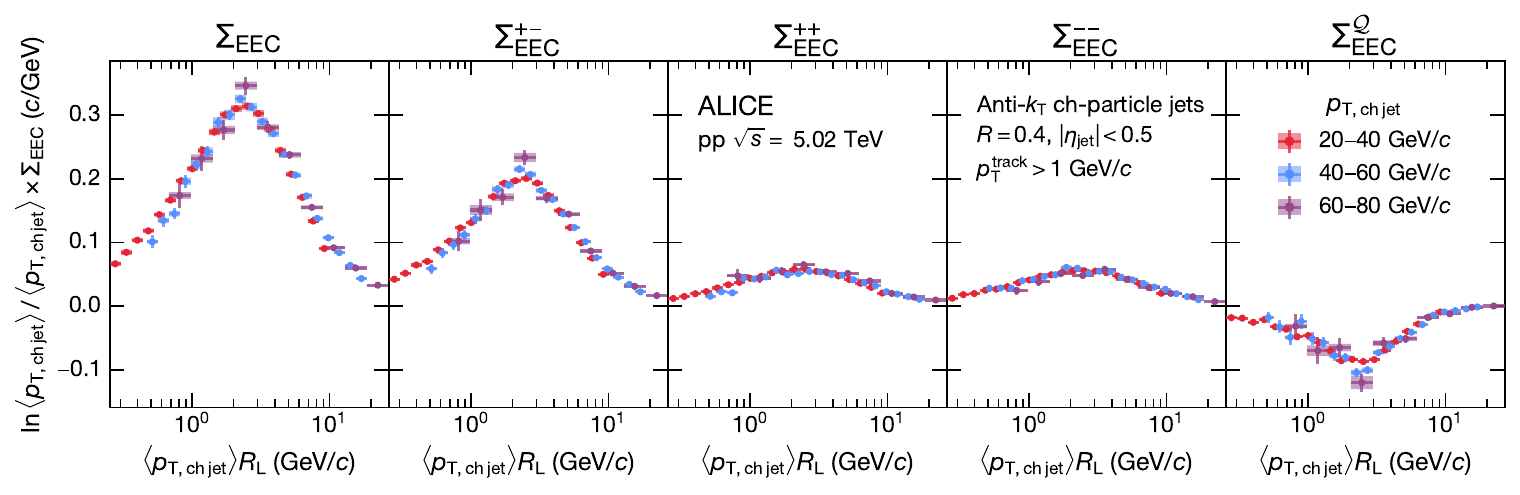}
\caption{The scaled inclusive and charged EECs in \pp collisions in each of the \ptchjet intervals.}
\label{fig:eec-pp-ptscale}
\end{figure}

Previous measurements of the inclusive EEC~\cite{aliceeecpp} demonstrated that the peak height is proportional to \ptchjetavg divided by the natural logarithm of \ptchjetavg,\footnote{In this text, \ptchjetavg is always normalized by 1 \GeVc to render it unitless before taking the logarithm.} and the angle at which the peak occurs is proportional to $1 / \ptchjetavg$. \ptchjetavg is the average \ptchjet in each interval of \ptchjet, and it was calculated in~\cite{aliceeecpp} from a \py simulation reweighted to match ALICE data. Thus, under appropriate scalings, the inclusive EEC collapses into a common curve~\cite{aliceeecpp}. \Cref{fig:eec-pp-ptscale} examines the \ptchjet dependence of the charged and inclusive EECs. The EECs in \cref{fig:eec-pp-ptscale} were scaled along the angular axis by \ptchjetavg, then scaled on the EEC axis by $\ln\ptchjetavg / \ptchjetavg$. \Cref{fig:eec-pp-ptscale} demonstrates that the charged EECs exhibit the same behavior as the inclusive EEC.

\begin{figure}[!htb]
\centering
\includegraphics[width=0.99\textwidth]{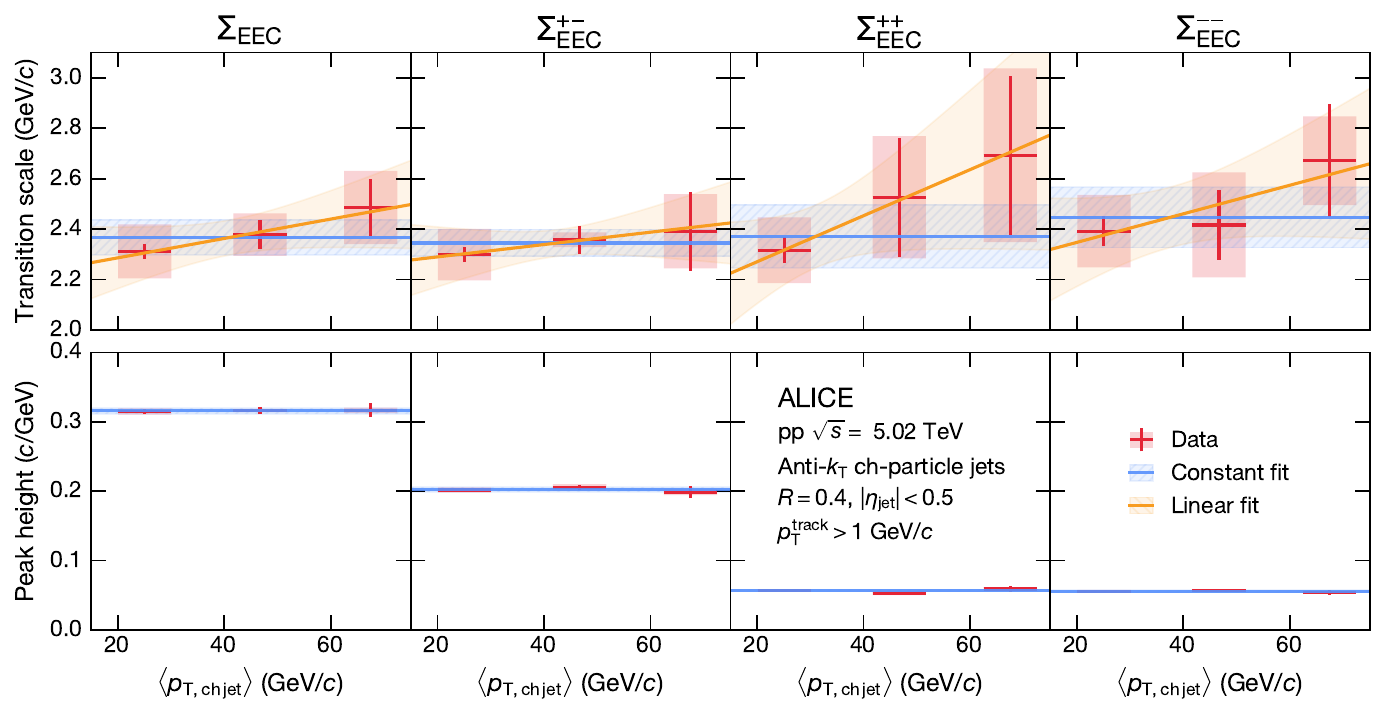}
\caption{The transition scales (top row) and peak heights (bottom row) of the scaled charged and inclusive EECs in ALICE data, as a function of the average \ptchjet in each interval. The statistical uncertainties are shown as vertical lines, and the systematic uncertainties are shown as shaded boxes. The horizontal uncertainties are set to $\pm 5$ \GeVc and are purely for visibility. Constant (blue) and linear fits (orange) across \ptchjetavg for each EEC are also shown as lines, with the surrounding shaded bands representing the 1$\sigma$ fit uncertainty.}
\label{fig:eec-pp-ptscale-comp}
\end{figure}%

A detailed characterization of this transition region is performed by fitting the scaled charge-selected and inclusive EECs near their peaks with a logarithmic Gaussian fit function, which was used in~\cite{aliceeecpp}: 
\begin{equation}
    \sigeec^\mathrm{fit} = C \exp\left( - \frac{(\ln (\ptchjetavg\RL) - \ln \mu)^2}{2\sigma^2} \right).
\end{equation}
The parameter $\mu$ represents the peak position along the $\ptchjetavg\RL$ axis (termed the transition scale in this text), while $C$ is the value of the fit function at $\ptchjetavg\RL = \mu$ (termed the peak height). The results for the transition scale and the peak height from the fits are shown as a function of \ptchjetavg in the top and bottom rows of \cref{fig:eec-pp-ptscale-comp}, respectively. Two sources of systematic uncertainty on the extracted fit parameters were evaluated: the choice of fit function, and the range of data points used in the fit. The extraction of the peak heights for each EEC are both highly precise (i.e. have small systematic and statistical uncertainty) and constant within uncertainties across the three intervals of \ptchjet. Constant fits to the transition scales for each of the charge-selected and inclusive EECs show that the transition scales are constant within uncertainties not only for the three \ptchjet intervals for each of the EECs, but also between the inclusive EEC and the three charge-selected EECs, occurring around 2.4 \GeVc. To study the observed increase in the central values of the transition scales across \ptchjetavg, linear fits were also performed. However, the resulting fits (orange lines in \cref{fig:eec-pp-ptscale-comp}) show significant overfitting within the uncertainty bounds, and the corresponding $1\sigma$ fit uncertainties are too large to rule out a constant fit.

\begin{figure}[!htb]
\centering
\includegraphics[width=0.99\textwidth]{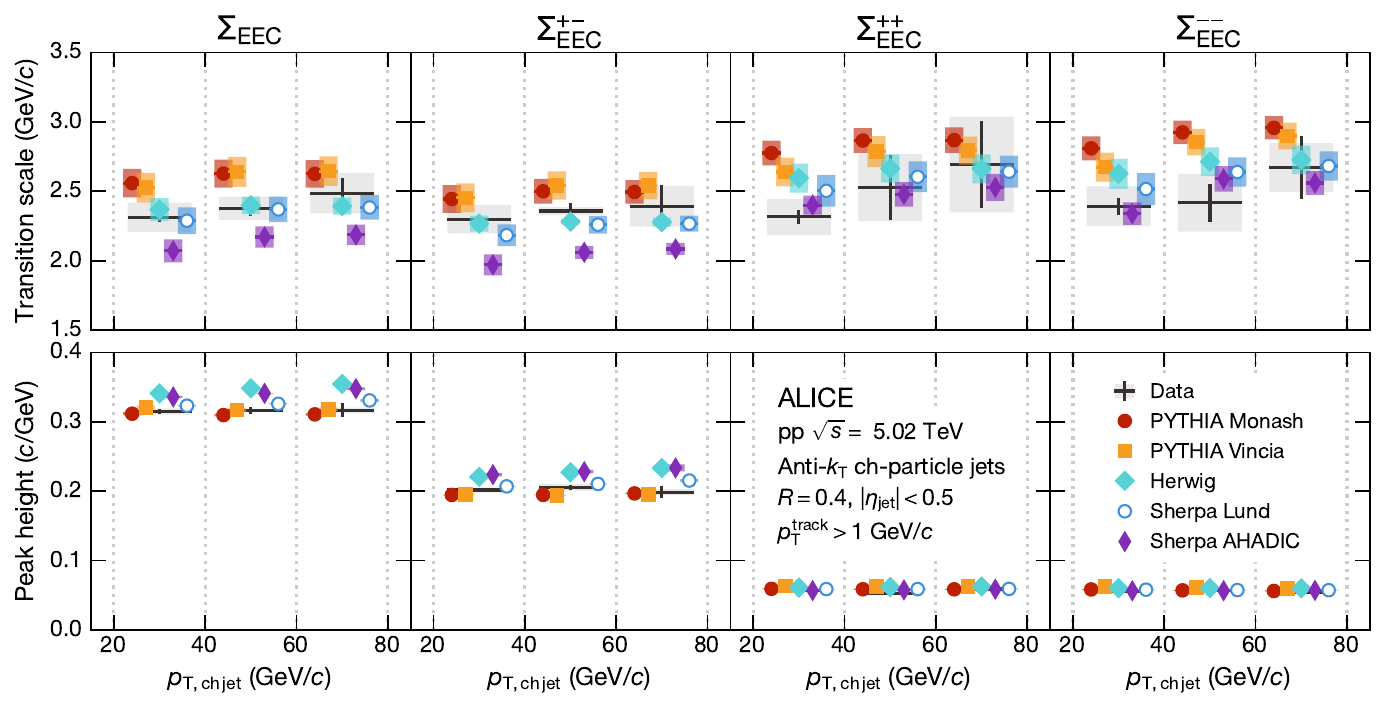}
\caption{The transition scales (top row) and peak heights (bottom row) in \pp collision data compared to MC models in each interval of \ptchjet. The data points for the MC models are offset within each interval for visibility only.}
\label{fig:eec-pp-ptscale-MC}
\end{figure}%

\Cref{fig:eec-pp-ptscale-MC} compares the \ptchjet dependence of the charged EEC peak in data to predictions from event generators. The same scaling and fitting procedure used for the data in \cref{fig:eec-pp-ptscale-comp} is applied to each event generator. In particular, the same values of \ptchjetavg that were used for the data are also used for all the event generators, in order to remove any dependence on the jet spectrum that might differ between models.

\sh Lund and AHADIC yield different transition scales and peak heights for the unlike-sign EEC, but similar values for the like-sign EECs. This conclusion is consistent with the observation in \cref{fig:eec-mc} that changes in hadronization modeling manifested primarily in changes in the unlike-sign EEC, but not in the like-sign EECs. On the contrary, changes in the overall EEC distribution do not necessarily lead to changes in the peak height and transition scale. Although \Cref{fig:eec-mc} shows that differences in shower modeling between \py Monash and Vincia modify the like-sign EEC, the corresponding peak heights and transition scales in \cref{fig:eec-pp-ptscale-MC} remain unchanged between the two models.

These two observations support the traditional interpretation that the \sigeec peak characterizes the dynamics of the hadronization transition~\cite{eecth,eec-rev}. However, we note that the peak of the charge-selected EECs cannot be interpreted in the same way, for two reasons. First, the relation between pair charge (i.e., like-sign or unlike-sign), which is defined at the hadron level, and the underlying QCD hadronization process is not clearly established. Second, the unlike-sign EEC transition scale is always at a lower value than either of the like-sign EECs within the same model. If the transition scales for the charge-selected EECs truly correspond to the energy scale at which hadronization occurs, then it is not obvious why like-sign and unlike-sign pairs would have different transition scales. It is currently not possible to conclude whether this hierarchy also exists within the data, due to the size of the uncertainties. However, all of the event generators studied in this analysis exhibit this behavior. It may be a general feature of all event generators, and would benefit from more precise future measurements to confirm or contradict the same pattern in data.

\subsection{Charged EEC asymmetry}

\begin{figure}[!htb]
\centering
\includegraphics[width=0.99\textwidth]{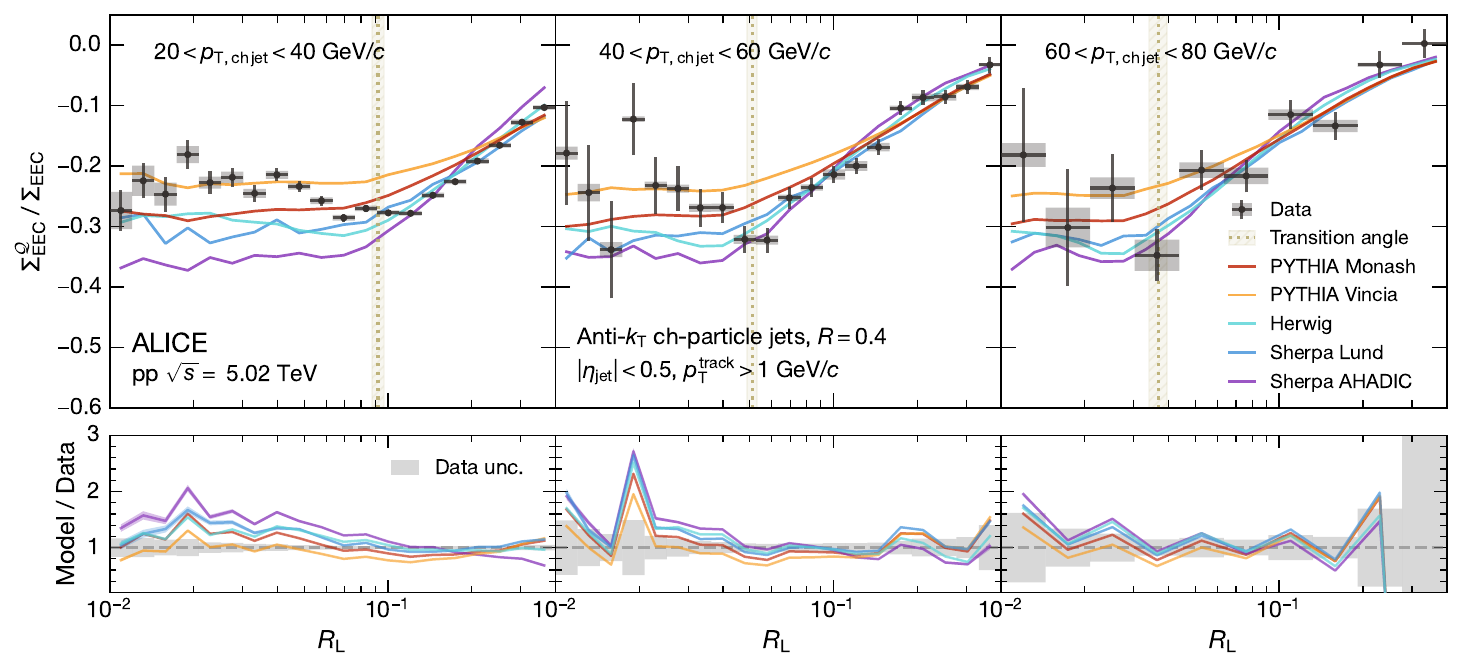}
\caption{The charged EEC asymmetry, defined as the ratio of the charge-weighted EEC to the inclusive EEC, in \pp collisions in the three \ptchjet intervals. The statistical uncertainties are shown as vertical lines, and the systematic uncertainties as shaded gray boxes. The transition angles, defined as the transition scale divided by \ptchjetavg in the given interval, are extracted from \cref{fig:eec-pp-ptscale} and displayed as vertical tan lines in the top row, with combined statistical and systematic uncertainties displayed as shaded bands. The relative combined statistical and systematic uncertainties on the data are shown as shaded boxes around the dashed unity line in the bottom row. Statistical uncertainties from each model are shown as shaded bands around the ratio lines, but are mostly too small to be visible.}
\label{fig:eec-pp-chratio}
\end{figure}%

Another way to characterize the charge and energy flow within jets is to measure not the absolute contributions of the individual charge-selected EECs to the inclusive EEC, as has been done so far, but rather their relative contributions. These relative contributions are quantified in the charged EEC asymmetry \sigasym, and compared to models in \cref{fig:eec-pp-chratio}. At large angles, the asymmetry is small (i.e. near zero), indicating near-equal contributions to the inclusive EEC from like- and unlike-sign pairs in the large-angle perturbative region. As the pair opening angle narrows, the relative unlike-sign contribution grows larger, and the asymmetry reaches its most negative value of approximately $-0.3$ for all \ptchjet intervals. The most negative asymmetry occurs at the transition angle, defined as the transition scale (which was shown in \cref{fig:eec-pp-ptscale-comp}) divided by \ptchjetavg for a given \ptchjet interval. The transition angle corresponds to the angle at which the EECs reach their maximal values in \cref{fig:eec-pp}. These transition angles are displayed for each \ptchjet interval in \cref{fig:eec-pp-chratio} as vertical tan lines.

Notably, past the transition scale, the asymmetry lessens slightly (i.e. becomes closer to zero) in the lowest \ptchjet interval before becoming consistent with a flat distribution, at around $-0.23$. The only model that qualitatively predicts this rebalancing of the asymmetry is \hw, but it is much less significant than that observed in the data. In all other generators, the asymmetry becomes flat at its most negative value as the angle decreases past the transition scale. A measurement by the STAR Collaboration~\cite{star} at $\sqrt{s} = 200$ GeV also observes the same rebalancing behavior, but to a much stronger degree: the asymmetries at small and large angles are nearly equal. Comparisons of the STAR data to \py 8 and \hw 7 show that these models also predict this rebalancing, but not to as strong a degree as in the STAR data. This may indicate that there are some missing or mismodeled dynamics for quark-initiated jets (which form a larger fraction of an inclusive jet sample at STAR center-of-mass energies compared to LHC energies) within these models, resulting in better quantitative descriptions of the ALICE data but better qualitative descriptions of the STAR data. It should also be noted that it is not yet possible to identify any meaningful rebalancing for jets with $40 < \ptchjet < 80$ \GeVc within the uncertainties, so more precise experimental measurements would benefit these studies.

As in \cref{fig:eec-mc}, no models can fully describe the measured charged EEC asymmetry, but some capture a subset. Only \hw fully captures the large angle behavior, and only \py Vincia describes the small angle behavior. \sh AHADIC tends to be significantly more asymmetric than the data.

\subsection{Charged EECs in \pPb collisions}

\begin{figure}[!htb]
\centering
\includegraphics[width=0.99\textwidth]{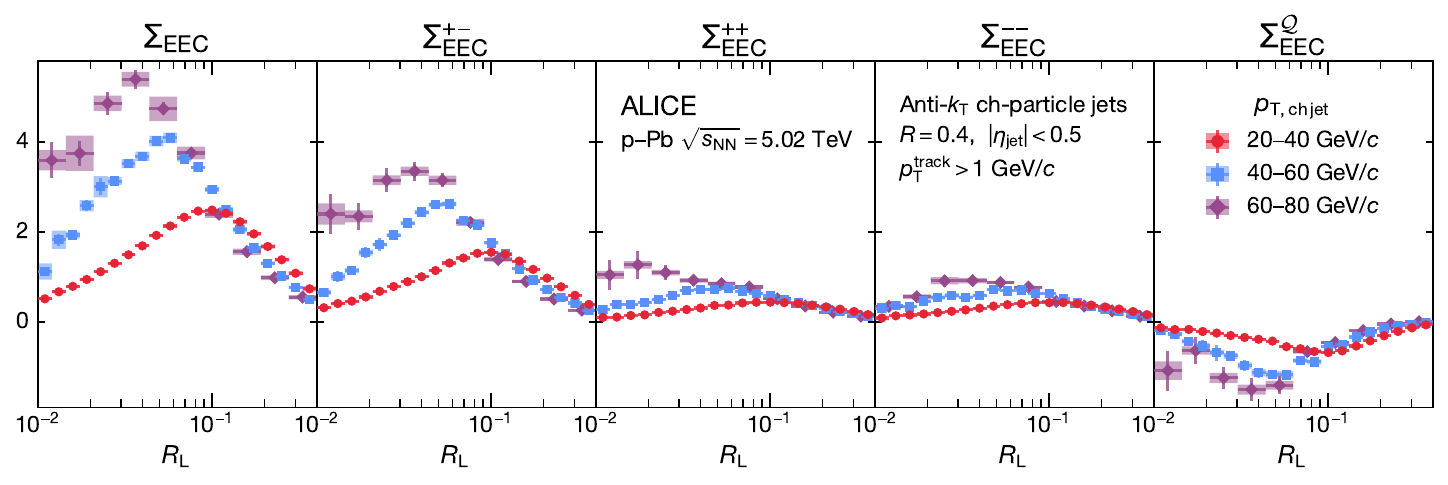}
\caption{The inclusive and charged EECs in \pPb collisions. Statistical uncertainties are shown as lines, and systematic uncertainties shown as shaded boxes.}
\label{fig:eec-ppb}
\end{figure}

\begin{figure}[!htb]
\centering
    \includegraphics[width=0.53\textwidth]{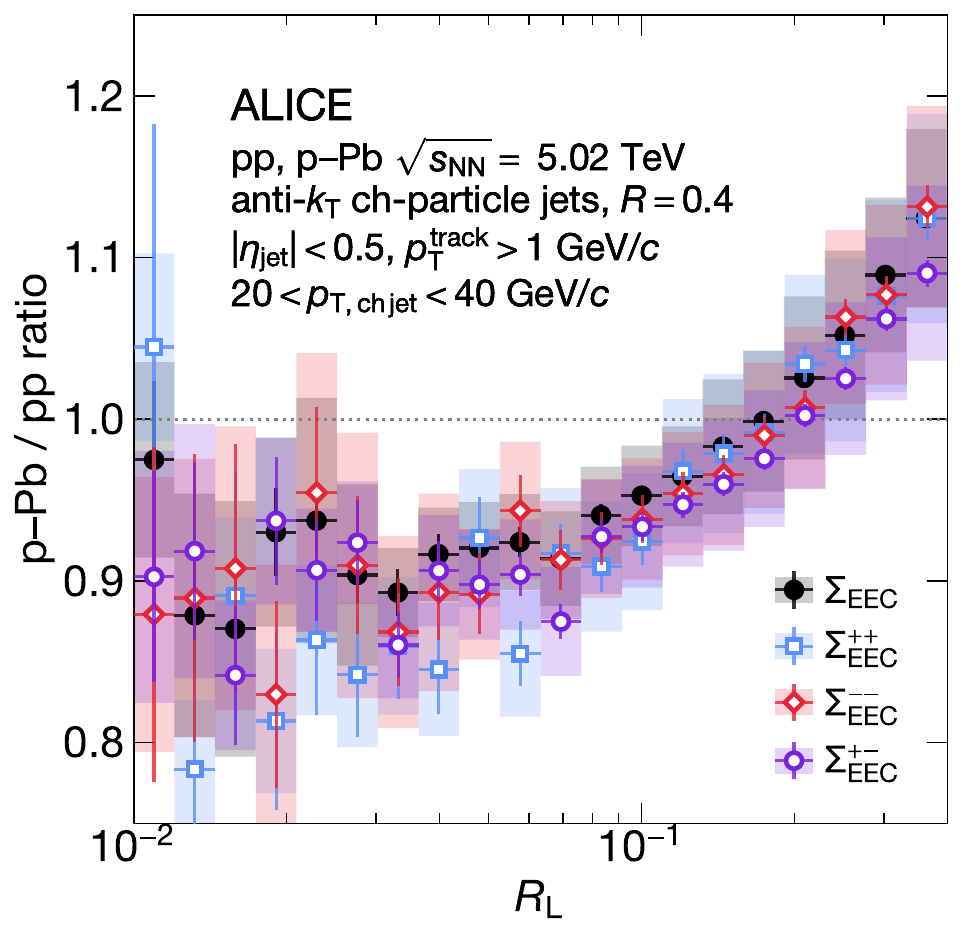}
    \caption{The ratio of charged and inclusive EECs in p--Pb to pp collisions for jets with $20 < \ptchjet < 40$ \GeVc.}
    \label{fig:eec-ratio}
\end{figure}

Previous ALICE EEC studies have indicated that cold nuclear matter in \pPb collisions modifies EECs, with a distinctive suppression below unity at small angles, which rises to an enhancement above unity at large angles in 20--40 \GeVc charged-particle jets~\cite{aliceppb}. The charged EECs probe whether this modification is dependent on charge. The inclusive and charged EECs in \pPb collisions for 20--40 \GeVc charged-particle jets, shown in \cref{fig:eec-ppb}, have the same general features as those from \pp collisions (see \cref{fig:eec-pp}). The ratio of the inclusive and charge-selected EECs in \pPb relative to \pp collisions is shown in \cref{fig:eec-ratio}. Within uncertainties, the modifications in the charge-selected EECs are identical to that in the inclusive EEC and no charge dependence is observed. Uncertainties at higher \ptchjet are too large to make concrete conclusions on charge dependence (or a lack thereof) and are not shown.

\subsection{Comparison to pQCD predictions}

New theoretical techniques have recently been developed to calculate infrared- and collinear-unsafe observables like the charged EECs in both elementary \pp collisions and those modified by nuclear matter, as in \pPb collisions. The charged EECs provide a valuable platform to validate these methods.

\begin{figure}[!htb]
\centering
\includegraphics[width=0.65\textwidth]{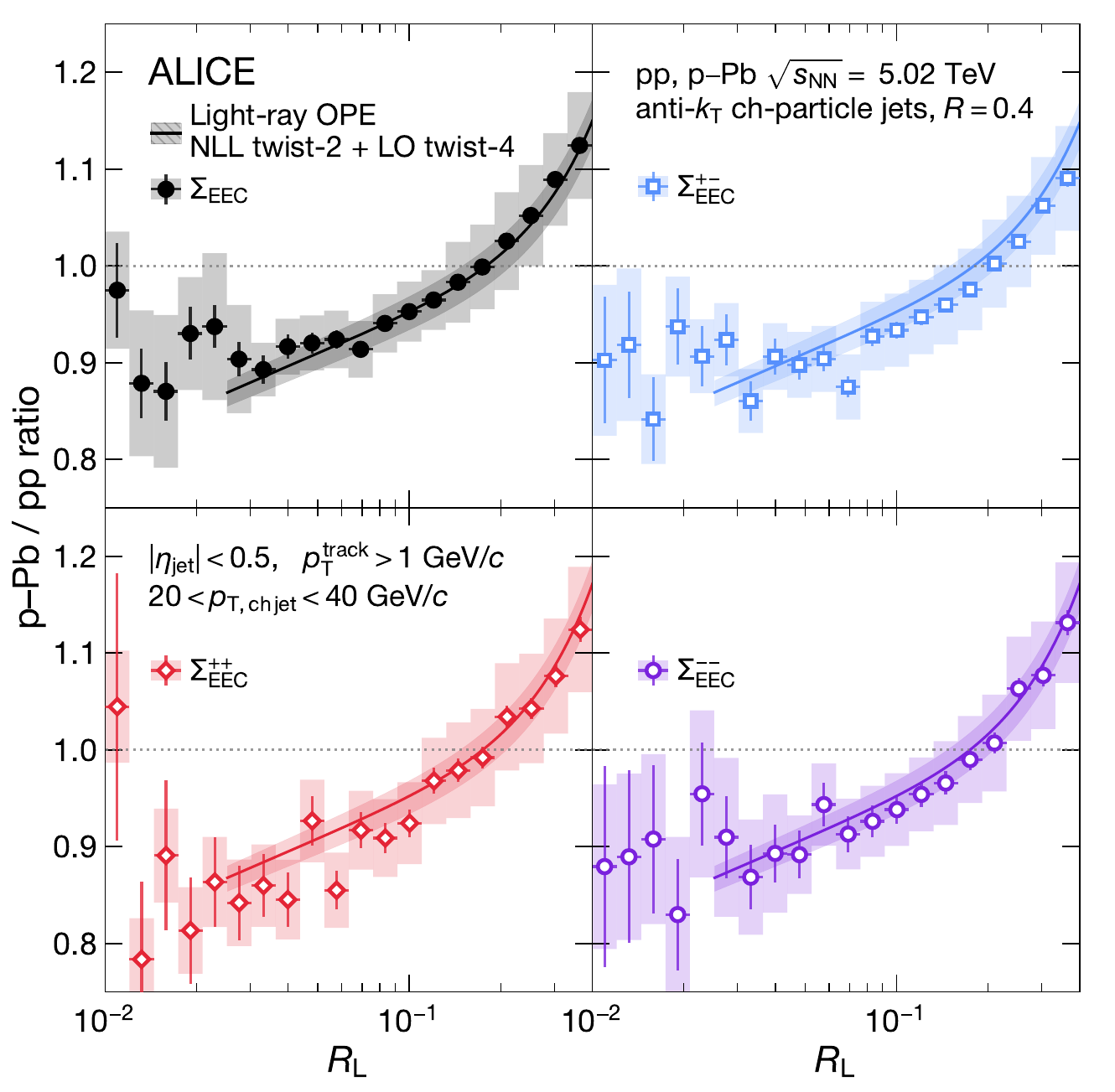}
\caption{Comparison of the charged and inclusive EEC ratio in \pPb to \pp collisions for 20--40 \GeVc charged-particle jets with a pQCD calculation based on~\cite{ceec_ratio_th}.}
\label{fig:eec-ratio-th}
\end{figure}%

\Cref{fig:eec-ratio-th} compares the charge-selected and inclusive EEC ratios measured in ALICE data to a pQCD calculation based on~\cite{ceec_ratio_th}. Based on a light-ray operator product expansion, the twist-2 component of this calculation, calculated to next-to-leading-log accuracy, incorporates local correlations that dominate the EEC in \pp collisions. The twist-4 matrix element, encoding nuclear matter effects, is estimated semi-classically, and is extracted from a fit to the inclusive EEC ratio in \pPb to \pp collisions presented in~\cite{aliceppb}. The excellent agreement between this calculation and the data shows that theoretical calculations can now account for cold nuclear matter effects. However, it should be noted that, because the twist-4 matrix element is fitted to the \sigeec ratio, the calculation and the measurement are not fully independent.

\section{Conclusions}
Measurements of charged energy--energy correlators in \pp and \pPb collisions are reported in this paper. The differential nature of the charged EECs provides insights into charge and energy flow dynamics that are hidden in inclusive EECs. Detailed comparisons to Monte Carlo event generators reveal that changes to jet evolution modeling may be isolated by charge-selected EECs. In particular, the like-sign EECs show sensitivity to modifications in shower modeling, and unlike-sign EECs are sensitive to different hadronization schemes. Models predict a hierarchy in the transition scale between different charge-selected EECs, with unlike-sign EECs peaking at smaller angles than like-sign EECs. However, the data are not yet precise enough to observe this behavior experimentally. The rebalancing of the charged EEC asymmetry at angles below the transition scale for low \ptchjet jets is entirely missing from model predictions, indicating incomplete dynamics in the models. In comparing charged EEC measurements in \pPb to \pp collisions, no significant charge dependence is observed within current uncertainties, and theoretical predictions are able to account for cold nuclear matter and charge in describing charged EECs.

%%%%%%%%%%%%%%%%%%%%%%%%%%%%%%%%
% end main text 
%%%%%%%%%%%%%%%%%%%%%%%%%%%%%%%%

%%%%% acknowledgements - handled by EB chairs 
\newenvironment{acknowledgement}{\relax}{\relax}
\begin{acknowledgement}
\section*{Acknowledgements}
% add specific acknowledgements here
% ...but please don't remove the line below: funding agencies
% will be acknowledged with a custom tex file handled by EB chairs after Collab Round 2
The ALICE Collaboration thanks C.~Andr\'es and J.~Holguin for providing theoretical predictions. 

% Version: 2026-06-02

The ALICE Collaboration would like to thank all its engineers and technicians for their invaluable contributions to the construction of the experiment and the CERN accelerator teams for the outstanding performance of the LHC complex.
The ALICE Collaboration gratefully acknowledges the resources and support provided by all Grid centres and the Worldwide LHC Computing Grid (WLCG) collaboration.
The ALICE Collaboration acknowledges the following funding agencies for their support in building and running the ALICE detector:
A. I. Alikhanyan National Science Laboratory (Yerevan Physics Institute) Foundation (ANSL), State Committee of Science and World Federation of Scientists (WFS), Armenia;
Austrian Academy of Sciences, Austrian Science Fund (FWF): [M 2467-N36] and Nationalstiftung f\"{u}r Forschung, Technologie und Entwicklung, Austria;
Ministry of Communications and High Technologies, National Nuclear Research Center, Azerbaijan;
Rede Nacional de Física de Altas Energias (Renafae), Financiadora de Estudos e Projetos (Finep), Funda\c{c}\~{a}o de Amparo \`{a} Pesquisa do Estado de S\~{a}o Paulo (FAPESP) and The Sao Paulo Research Foundation  (FAPESP), Brazil;
Bulgarian Ministry of Education and Science, within the National Roadmap for Research Infrastructures 2020-2027 (object CERN), Bulgaria;
Ministry of Education of China (MOEC) , Ministry of Science \& Technology of China (MSTC) and National Natural Science Foundation of China (NSFC), China;
Ministry of Science and Education and Croatian Science Foundation, Croatia;
Centro de Aplicaciones Tecnol\'{o}gicas y Desarrollo Nuclear (CEADEN), Cubaenerg\'{\i}a, Cuba;
Ministry of Education, Youth and Sports of the Czech Republic, Czech Republic;
The Danish Council for Independent Research | Natural Sciences, the VILLUM FONDEN and Danish National Research Foundation (DNRF), Denmark;
Helsinki Institute of Physics (HIP), Finland;
Commissariat \`{a} l'Energie Atomique (CEA) and Institut National de Physique Nucl\'{e}aire et de Physique des Particules (IN2P3) and Centre National de la Recherche Scientifique (CNRS), France;
Bundesministerium f\"{u}r Forschung, Technologie und Raumfahrt (BMFTR) and GSI Helmholtzzentrum f\"{u}r Schwerionenforschung GmbH, Germany;
National Research, Development and Innovation Office, Hungary;
Department of Atomic Energy Government of India (DAE), Department of Science and Technology, Government of India (DST), University Grants Commission, Government of India (UGC) and Council of Scientific and Industrial Research (CSIR), India;
National Research and Innovation Agency - BRIN, Indonesia;
Istituto Nazionale di Fisica Nucleare (INFN), Italy;
Japanese Ministry of Education, Culture, Sports, Science and Technology (MEXT) and Japan Society for the Promotion of Science (JSPS) KAKENHI, Japan;
Consejo Nacional de Ciencia (CONACYT) y Tecnolog\'{i}a, through Fondo de Cooperaci\'{o}n Internacional en Ciencia y Tecnolog\'{i}a (FONCICYT) and Direcci\'{o}n General de Asuntos del Personal Academico (DGAPA), Mexico;
Nederlandse Organisatie voor Wetenschappelijk Onderzoek (NWO), Netherlands;
The Research Council of Norway, Norway;
Pontificia Universidad Cat\'{o}lica del Per\'{u}, Peru;
Ministry of Science and Higher Education, National Science Centre and WUT ID-UB, Poland;
Korea Institute of Science and Technology Information and National Research Foundation of Korea (NRF), Republic of Korea;
Ministry of Education and Scientific Research, Institute of Atomic Physics, Ministry of Research and Innovation and Institute of Atomic Physics and Universitatea Nationala de Stiinta si Tehnologie Politehnica Bucuresti, Romania;
Ministerstvo skolstva, vyskumu, vyvoja a mladeze SR, Slovakia;
National Research Foundation of South Africa, South Africa;
Swedish Research Council (VR) and Knut \& Alice Wallenberg Foundation (KAW), Sweden;
European Organization for Nuclear Research, Switzerland;
Suranaree University of Technology (SUT), National Science and Technology Development Agency (NSTDA) and National Science, Research and Innovation Fund (NSRF via PMU-B B05F650021), Thailand;
Turkish Energy, Nuclear and Mineral Research Agency (TENMAK), Turkey;
National Academy of  Sciences of Ukraine, Ukraine;
Science and Technology Facilities Council (STFC), United Kingdom;
National Science Foundation of the United States of America (NSF) and United States Department of Energy, Office of Nuclear Physics (DOE NP), United States of America.
In addition, individual groups or members have received support from:
Czech Science Foundation (grant no. 23-07499S), Czech Republic;
FORTE project, reg.\ no.\ CZ.02.01.01/00/22\_008/0004632, Czech Republic, co-funded by the European Union, Czech Republic;
European Research Council (grant no. 101220549), European Union;
Deutsche Forschungs Gemeinschaft (DFG, German Research Foundation) ``Neutrinos and Dark Matter in Astro- and Particle Physics'' (grant no. SFB 1258), Germany;
CONVECS project, CUP C97H23001700002 FESR 2021-2027 program, Italy.

\end{acknowledgement}

%%%%%%%% Bibliography
\bibliographystyle{utphys}   % Remember we use title in the biblio
\bibliography{bibliography}
%\input {bibliography.tex}  

%%%%%%%%%%%%%%%%%%%%%%%%%%%%%%%%
% Appendices: yours (if any) + authorlist
%%%%%%%%%%%%%%%%%%%%%%%%%%%%%%%%
\newpage
\appendix

%
%\input{} % put your appendices here (if any)
%

%%%%% Authorlist - please do not touch: handled by EB chairs 
\section{The ALICE Collaboration}
\label{app:collab}
% ALICE Collaboration author list for 2026-06-02
\begin{flushleft} 
\small

D.A.H.~Abdallah\,\orcidlink{0000-0003-4768-2718}\,$^{\rm 134}$, 
I.J.~Abualrob\,\orcidlink{0009-0005-3519-5631}\,$^{\rm 112}$, 
S.~Acharya\,\orcidlink{0000-0002-9213-5329}\,$^{\rm 49}$, 
K.~Agarwal\,\orcidlink{0000-0001-5781-3393}\,$^{\rm II,}$$^{\rm 23}$, 
G.~Aglieri Rinella\,\orcidlink{0000-0002-9611-3696}\,$^{\rm 32}$, 
L.~Aglietta\,\orcidlink{0009-0003-0763-6802}\,$^{\rm 24}$, 
N.~Agrawal\,\orcidlink{0000-0003-0348-9836}\,$^{\rm 25}$, 
Z.~Ahammed\,\orcidlink{0000-0001-5241-7412}\,$^{\rm 132}$, 
S.~Ahmad\,\orcidlink{0000-0003-0497-5705}\,$^{\rm 15}$, 
Z.~Akbar$^{\rm 79}$
% ...

\section*{Affiliation Notes}

$^{\rm I}$ Deceased\\
$^{\rm II}$ Also at: INFN Trieste\\
$^{\rm III}$ Also at: Fondazione Bruno Kessler (FBK), Trento, Italy\\
$^{\rm IV}$ Also at: Czech Technical University in Prague (CZ)\\
$^{\rm V}$ Also at: Instituto de Fisica da Universidade de Sao Paulo\\
$^{\rm VI}$ Also at: Dipartimento DET del Politecnico di Torino, Turin, Italy\\
$^{\rm VII}$ Also at: Department of Applied Physics, Aligarh Muslim University, Aligarh, India\\
$^{\rm VIII}$ Also at: Institute of Theoretical Physics, University of Wroclaw, Poland\\
$^{\rm IX}$ Also at: Facultad de Ciencias, Universidad Nacional Aut\'{o}noma de M\'{e}xico, Mexico City, Mexico\\

\section*{Collaboration Institutes}

$^{1}$ A.I. Alikhanyan National Science Laboratory (Yerevan Physics Institute) Foundation, Yerevan, Armenia\\
$^{2}$ AGH University of Krakow, Cracow, Poland\\
$^{3}$ Bogolyubov Institute for Theoretical Physics, National Academy of Sciences of Ukraine, Kyiv, Ukraine\\
$^{4}$ Bose Institute, Department of Physics  and Centre for Astroparticle Physics and Space Science (CAPSS), Kolkata, India\\
$^{5}$ California Polytechnic State University, San Luis Obispo, California, United States\\
$^{6}$ Central China Normal University, Wuhan, China\\
$^{7}$ Centro de Aplicaciones Tecnol\'{o}gicas y Desarrollo Nuclear (CEADEN), Havana, Cuba\\
$^{8}$ Centro de Investigaci\'{o}n y de Estudios Avanzados (CINVESTAV), Mexico City and M\'{e}rida, Mexico\\
$^{9}$ Chicago State University, Chicago, Illinois, United States\\
$^{10}$ China Nuclear Data Center, China Institute of Atomic Energy, Beijing, China\\
$^{11}$ China University of Geosciences, Wuhan, China\\
$^{12}$ Chungbuk National University, Cheongju, Republic of Korea\\
$^{13}$ Comenius University Bratislava, Faculty of Mathematics, Physics and Informatics, Bratislava, Slovak Republic\\
$^{14}$ Creighton University, Omaha, Nebraska, United States\\
$^{15}$ Department of Physics, Aligarh Muslim University, Aligarh, India\\
$^{16}$ Department of Physics, Pusan National University, Pusan, Republic of Korea\\
$^{17}$ Department of Physics, Sejong University, Seoul, Republic of Korea\\
$^{18}$ Department of Physics, University of California, Berkeley, California, United States\\
$^{19}$ Department of Physics, University of Oslo, Oslo, Norway\\
$^{20}$ Department of Physics and Technology, University of Bergen, Bergen, Norway\\
$^{21}$ Dipartimento di Fisica, Universit\`{a} di Pavia, Pavia, Italy\\
$^{22}$ Dipartimento di Fisica dell'Universit\`{a} and Sezione INFN, Cagliari, Italy\\
$^{23}$ Dipartimento di Fisica dell'Universit\`{a} and Sezione INFN, Trieste, Italy\\
$^{24}$ Dipartimento di Fisica dell'Universit\`{a} and Sezione INFN, Turin, Italy\\
$^{25}$ Dipartimento di Fisica e Astronomia dell'Universit\`{a} and Sezione INFN, Bologna, Italy\\
$^{26}$ Dipartimento di Fisica e Astronomia dell'Universit\`{a} and Sezione INFN, Catania, Italy\\
$^{27}$ Dipartimento di Fisica e Astronomia dell'Universit\`{a} and Sezione INFN, Padova, Italy\\
$^{28}$ Dipartimento di Fisica `E.R.~Caianiello' dell'Universit\`{a} and Gruppo Collegato INFN, Salerno, Italy\\
$^{29}$ Dipartimento DISAT del Politecnico and Sezione INFN, Turin, Italy\\
$^{30}$ Dipartimento di Scienze MIFT, Universit\`{a} di Messina, Messina, Italy\\
$^{31}$ Dipartimento Interateneo di Fisica `M.~Merlin' and Sezione INFN, Bari, Italy\\
$^{32}$ European Organization for Nuclear Research (CERN), Geneva, Switzerland\\
$^{33}$ Faculty of Electrical Engineering, Mechanical Engineering and Naval Architecture, University of Split, Split, Croatia\\
$^{34}$ Faculty of Nuclear Sciences and Physical Engineering, Czech Technical University in Prague, Prague, Czech Republic\\
$^{35}$ Faculty of Physics, Sofia University, Sofia, Bulgaria\\
$^{36}$ Faculty of Science, P.J.~\v{S}af\'{a}rik University, Ko\v{s}ice, Slovak Republic\\
$^{37}$ Faculty of Technology, Environmental and Social Sciences, Bergen, Norway\\
$^{38}$ Frankfurt Institute for Advanced Studies, Johann Wolfgang Goethe-Universit\"{a}t Frankfurt, Frankfurt, Germany\\
$^{39}$ Fudan University, Shanghai, China\\
$^{40}$ Gauhati University, Department of Physics, Guwahati, India\\
$^{41}$ Helmholtz-Institut f\"{u}r Strahlen- und Kernphysik, Rheinische Friedrich-Wilhelms-Universit\"{a}t Bonn, Bonn, Germany\\
$^{42}$ Helsinki Institute of Physics (HIP), Helsinki, Finland\\
$^{43}$ High Energy Physics Group,  Universidad Aut\'{o}noma de Puebla, Puebla, Mexico\\
$^{44}$ Horia Hulubei National Institute of Physics and Nuclear Engineering, Bucharest, Romania\\
$^{45}$ HUN-REN Wigner Research Centre for Physics, Budapest, Hungary\\
$^{46}$ Indian Institute of Technology Bombay (IIT), Mumbai, India\\
$^{47}$ Indian Institute of Technology Indore, Indore, India\\
$^{48}$ INFN, Laboratori Nazionali di Frascati, Frascati, Italy\\
$^{49}$ INFN, Sezione di Bari, Bari, Italy\\
$^{50}$ INFN, Sezione di Bologna, Bologna, Italy\\
$^{51}$ INFN, Sezione di Cagliari, Cagliari, Italy\\
$^{52}$ INFN, Sezione di Catania, Catania, Italy\\
$^{53}$ INFN, Sezione di Padova, Padova, Italy\\
$^{54}$ INFN, Sezione di Pavia, Pavia, Italy\\
$^{55}$ INFN, Sezione di Torino, Turin, Italy\\
$^{56}$ INFN, Sezione di Trieste, Trieste, Italy\\
$^{57}$ Inha University, Incheon, Republic of Korea\\
$^{58}$ Institute for Gravitational and Subatomic Physics (GRASP), Utrecht University/Nikhef, Utrecht, Netherlands\\
$^{59}$ Institute of Experimental Physics, Slovak Academy of Sciences, Ko\v{s}ice, Slovak Republic\\
$^{60}$ Institute of Physics, Homi Bhabha National Institute, Bhubaneswar, India\\
$^{61}$ Institute of Physics of the Czech Academy of Sciences, Prague, Czech Republic\\
$^{62}$ Institute of Space Science (ISS), Bucharest, Romania\\
$^{63}$ Institut f\"{u}r Kernphysik, Johann Wolfgang Goethe-Universit\"{a}t Frankfurt, Frankfurt, Germany\\
$^{64}$ Instituto de Ciencias Nucleares, Universidad Nacional Aut\'{o}noma de M\'{e}xico, Mexico City, Mexico\\
$^{65}$ Instituto de F\'{i}sica, Universidade Federal do Rio Grande do Sul (UFRGS), Porto Alegre, Brazil\\
$^{66}$ Instituto de F\'{\i}sica, Universidad Nacional Aut\'{o}noma de M\'{e}xico, Mexico City, Mexico\\
$^{67}$ iThemba LABS, National Research Foundation, Somerset West, South Africa\\
$^{68}$ Jeonbuk National University, Jeonju, Republic of Korea\\
$^{69}$ Korea Institute of Science and Technology Information, Daejeon, Republic of Korea\\
$^{70}$ Laboratoire de Physique Subatomique et de Cosmologie, Universit\'{e} Grenoble-Alpes, CNRS-IN2P3, Grenoble, France\\
$^{71}$ Lawrence Berkeley National Laboratory, Berkeley, California, United States\\
$^{72}$ Lund University Department of Physics, Division of Particle Physics, Lund, Sweden\\
$^{73}$ Marietta Blau Institute, Vienna, Austria\\
$^{74}$ Nagasaki Institute of Applied Science, Nagasaki, Japan\\
$^{75}$ Nara Women{'}s University (NWU), Nara, Japan\\
$^{76}$ National Centre for Nuclear Research, Warsaw, Poland\\
$^{77}$ National Institute of Science Education and Research, Homi Bhabha National Institute, Jatni, India\\
$^{78}$ National Nuclear Research Center, Baku, Azerbaijan\\
$^{79}$ National Research and Innovation Agency - BRIN, Jakarta, Indonesia\\
$^{80}$ Niels Bohr Institute, University of Copenhagen, Copenhagen, Denmark\\
$^{81}$ Nikhef, National institute for subatomic physics, Amsterdam, Netherlands\\
$^{82}$ Nuclear Physics Group, STFC Daresbury Laboratory, Daresbury, United Kingdom\\
$^{83}$ Nuclear Physics Institute of the Czech Academy of Sciences, Husinec-\v{R}e\v{z}, Czech Republic\\
$^{84}$ Oak Ridge National Laboratory, Oak Ridge, Tennessee, United States\\
$^{85}$ Ohio State University, Columbus, Ohio, United States\\
$^{86}$ Physics department, Faculty of science, University of Zagreb, Zagreb, Croatia\\
$^{87}$ Physics Department, Panjab University, Chandigarh, India\\
$^{88}$ Physics Department, University of Jammu, Jammu, India\\
$^{89}$ Physics Program and International Institute for Sustainability with Knotted Chiral Meta Matter (WPI-SKCM$^{2}$), Hiroshima University, Hiroshima, Japan\\
$^{90}$ Physikalisches Institut, Eberhard-Karls-Universit\"{a}t T\"{u}bingen, T\"{u}bingen, Germany\\
$^{91}$ Physikalisches Institut, Ruprecht-Karls-Universit\"{a}t Heidelberg, Heidelberg, Germany\\
$^{92}$ Physik Department, Technische Universit\"{a}t M\"{u}nchen, Munich, Germany\\
$^{93}$ Politecnico di Bari and Sezione INFN, Bari, Italy\\
$^{94}$ Research Division and ExtreMe Matter Institute EMMI, GSI Helmholtzzentrum f\"ur Schwerionenforschung GmbH, Darmstadt, Germany\\
$^{95}$ Saga University, Saga, Japan\\
$^{96}$ Saha Institute of Nuclear Physics, Homi Bhabha National Institute, Kolkata, India\\
$^{97}$ School of Physics and Astronomy, University of Birmingham, Birmingham, United Kingdom\\
$^{98}$ Secci\'{o}n F\'{\i}sica, Departamento de Ciencias, Pontificia Universidad Cat\'{o}lica del Per\'{u}, Lima, Peru\\
$^{99}$ SUBATECH, IMT Atlantique, Nantes Universit\'{e}, CNRS-IN2P3, Nantes, France\\
$^{100}$ Sungkyunkwan University, Suwon City, Republic of Korea\\
$^{101}$ Suranaree University of Technology, Nakhon Ratchasima, Thailand\\
$^{102}$ Technical University of Ko\v{s}ice, Ko\v{s}ice, Slovak Republic\\
$^{103}$ The Henryk Niewodniczanski Institute of Nuclear Physics, Polish Academy of Sciences, Cracow, Poland\\
$^{104}$ The University of Texas at Austin, Austin, Texas, United States\\
$^{105}$ Universidad Aut\'{o}noma de Sinaloa, Culiac\'{a}n, Mexico\\
$^{106}$ Universidade de S\~{a}o Paulo (USP), S\~{a}o Paulo, Brazil\\
$^{107}$ Universidade Estadual de Campinas (UNICAMP), Campinas, Brazil\\
$^{108}$ Universidade Federal do ABC, Santo Andre, Brazil\\
$^{109}$ Universitatea Nationala de Stiinta si Tehnologie Politehnica Bucuresti, Bucharest, Romania\\
$^{110}$ University of Cape Town, Cape Town, South Africa\\
$^{111}$ University of Derby, Derby, United Kingdom\\
$^{112}$ University of Houston, Houston, Texas, United States\\
$^{113}$ University of Jyv\"{a}skyl\"{a}, Jyv\"{a}skyl\"{a}, Finland\\
$^{114}$ University of Kansas, Lawrence, Kansas, United States\\
$^{115}$ University of Liverpool, Liverpool, United Kingdom\\
$^{116}$ University of Science and Technology of China, Hefei, China\\
$^{117}$ University of Silesia in Katowice, Katowice, Poland\\
$^{118}$ University of South-Eastern Norway, Kongsberg, Norway\\
$^{119}$ University of Tennessee, Knoxville, Tennessee, United States\\
$^{120}$ University of the Witwatersrand, Johannesburg, South Africa\\
$^{121}$ University of Tokyo, Tokyo, Japan\\
$^{122}$ University of Tsukuba, Tsukuba, Japan\\
$^{123}$ Universit\"{a}t M\"{u}nster, Institut f\"{u}r Kernphysik, M\"{u}nster, Germany\\
$^{124}$ Universit\'{e} Clermont Auvergne, CNRS/IN2P3, LPC, Clermont-Ferrand, France\\
$^{125}$ Universit\'{e} de Lyon, CNRS/IN2P3, Institut de Physique des 2 Infinis de Lyon, Lyon, France\\
$^{126}$ Universit\'{e} de Strasbourg, CNRS, IPHC UMR 7178, F-67000 Strasbourg, France, Strasbourg, France\\
$^{127}$ Universit\'{e} Paris-Saclay, Centre d'Etudes de Saclay (CEA), IRFU, D\'{e}partment de Physique Nucl\'{e}aire (DPhN), Saclay, France\\
$^{128}$ Universit\'{e}  Paris-Saclay, CNRS/IN2P3, IJCLab, Orsay, France\\
$^{129}$ Universit\`{a} degli Studi di Foggia, Foggia, Italy\\
$^{130}$ Universit\`{a} del Piemonte Orientale, Vercelli, Italy\\
$^{131}$ Universit\`{a} di Brescia, Brescia, Italy\\
$^{132}$ Variable Energy Cyclotron Centre, Homi Bhabha National Institute, Kolkata, India\\
$^{133}$ Warsaw University of Technology, Warsaw, Poland\\
$^{134}$ Wayne State University, Detroit, Michigan, United States\\
$^{135}$ Yale University, New Haven, Connecticut, United States\\
$^{136}$ Yildiz Technical University, Istanbul, Turkey\\
$^{137}$ Yonsei University, Seoul, Republic of Korea\\
$^{138}$ Affiliated with an institute formerly covered by a cooperation agreement with CERN\\
$^{139}$ Affiliated with an international laboratory covered by a cooperation agreement with CERN.\\

\end{flushleft} 
  
\end{document}